\documentclass[reprint,amsmath,amssymb,aps,floatfix,superscriptaddress,showpacs,prl]{revtex4-1}

\usepackage{graphicx}
\usepackage{dcolumn}
\usepackage{bm}
\usepackage{color}
\usepackage{subfigure}
\usepackage{url}

\begin{document}
{

\title{Direct Measurement of the Cosmic-Ray Carbon and Oxygen Spectra \\from 10 GeV/$n$ to 2.2 TeV/$n$
  with the Calorimetric Electron Telescope \\on the International Space Station}

\author{O.~Adriani}
\affiliation{Department of Physics, University of Florence, Via Sansone, 1 - 50019 Sesto, Fiorentino, Italy}
\affiliation{INFN Sezione di Florence, Via Sansone, 1 - 50019 Sesto, Fiorentino, Italy}
\author{Y.~Akaike}
 \email{yakaike@aoni.waseda.jp}
\affiliation{Waseda Research Institute for Science and Engineering, Waseda University, 17 Kikuicho,  Shinjuku, Tokyo 162-0044, Japan}
\affiliation{JEM Utilization Center, Human Spaceflight Technology Directorate, Japan Aerospace Exploration Agency, 2-1-1 Sengen, Tsukuba, Ibaraki 305-8505, Japan}
\author{K.~Asano}
\affiliation{Institute for Cosmic Ray Research, The University of Tokyo, 5-1-5 Kashiwa-no-Ha, Kashiwa, Chiba 277-8582, Japan}
\author{Y.~Asaoka}
\affiliation{Institute for Cosmic Ray Research, The University of Tokyo, 5-1-5 Kashiwa-no-Ha, Kashiwa, Chiba 277-8582, Japan}
\author{M.G.~Bagliesi}
\affiliation{Department of Physical Sciences, Earth and Environment, University of Siena, via Roma 56, 53100 Siena, Italy}
\affiliation{INFN Sezione di Pisa, Polo Fibonacci, Largo B. Pontecorvo, 3 - 56127 Pisa, Italy}
\author{E.~Berti} 
\affiliation{Department of Physics, University of Florence, Via Sansone, 1 - 50019 Sesto, Fiorentino, Italy}
\affiliation{INFN Sezione di Florence, Via Sansone, 1 - 50019 Sesto, Fiorentino, Italy}
\author{G.~Bigongiari}
\affiliation{Department of Physical Sciences, Earth and Environment, University of Siena, via Roma 56, 53100 Siena, Italy}
\affiliation{INFN Sezione di Pisa, Polo Fibonacci, Largo B. Pontecorvo, 3 - 56127 Pisa, Italy}
\author{W.R.~Binns}
\affiliation{Department of Physics and McDonnell Center for the Space Sciences, Washington University, One Brookings Drive, St. Louis, MO 63130-4899, USA}
\author{M.~Bongi}
\affiliation{Department of Physics, University of Florence, Via Sansone, 1 - 50019 Sesto, Fiorentino, Italy}
\affiliation{INFN Sezione di Florence, Via Sansone, 1 - 50019 Sesto, Fiorentino, Italy}
\author{P.~Brogi}
\affiliation{Department of Physical Sciences, Earth and Environment, University of Siena, via Roma 56, 53100 Siena, Italy}
\affiliation{INFN Sezione di Pisa, Polo Fibonacci, Largo B. Pontecorvo, 3 - 56127 Pisa, Italy}
\author{A.~Bruno}
\affiliation{Heliospheric Physics Laboratory, NASA/GSFC, Greenbelt, Maryland 20771, USA}
\author{J.H.~Buckley}
\affiliation{Department of Physics and McDonnell Center for the Space Sciences, Washington University, One Brookings Drive, St. Louis, MO 63130-4899, USA}
\author{N.~Cannady}
\affiliation{Center for Space Sciences and Technology, University of Maryland, Baltimore County, 1000 Hilltop Circle, Baltimore, Maryland 21250, USA}
\affiliation{Astroparticle Physics Laboratory, NASA/GSFC, Greenbelt, Maryland 20771, USA}
\affiliation{Center for Research and Exploration in Space Sciences and Technology, NASA/GSFC, Greenbelt, Maryland 20771, USA}
\author{G.~Castellini}
\affiliation{Institute of Applied Physics (IFAC),  National Research Council (CNR), Via Madonna del Piano, 10, 50019 Sesto, Fiorentino, Italy}
\author{C.~Checchia}
\affiliation{Department of Physics, University of Florence, Via Sansone, 1 - 50019 Sesto, Fiorentino, Italy}
\affiliation{INFN Sezione di Florence, Via Sansone, 1 - 50019 Sesto, Fiorentino, Italy}
\author{M.L.~Cherry}
\affiliation{Department of Physics and Astronomy, Louisiana State University, 202 Nicholson Hall, Baton Rouge, LA 70803, USA}
\author{G.~Collazuol}
\affiliation{Department of Physics and Astronomy, University of Padova, Via Marzolo, 8, 35131 Padova, Italy}
\affiliation{INFN Sezione di Padova, Via Marzolo, 8, 35131 Padova, Italy} 
\author{K.~Ebisawa}
\affiliation{Institute of Space and Astronautical Science, Japan Aerospace Exploration Agency, 3-1-1 Yoshinodai, Chuo, Sagamihara, Kanagawa 252-5210, Japan}
\author{H.~Fuke}
\affiliation{Institute of Space and Astronautical Science, Japan Aerospace Exploration Agency, 3-1-1 Yoshinodai, Chuo, Sagamihara, Kanagawa 252-5210, Japan}
\author{S.~Gonzi}
\affiliation{Department of Physics, University of Florence, Via Sansone, 1 - 50019 Sesto, Fiorentino, Italy}
\affiliation{INFN Sezione di Florence, Via Sansone, 1 - 50019 Sesto, Fiorentino, Italy}
\author{T.G.~Guzik}
\affiliation{Department of Physics and Astronomy, Louisiana State University, 202 Nicholson Hall, Baton Rouge, LA 70803, USA}
\author{T.~Hams}
\affiliation{Center for Space Sciences and Technology, University of Maryland, Baltimore County, 1000 Hilltop Circle, Baltimore, Maryland 21250, USA}
\author{K.~Hibino}
\affiliation{Kanagawa University, 3-27-1 Rokkakubashi, Kanagawa, Yokohama, Kanagawa 221-8686, Japan}
\author{M.~Ichimura}
\affiliation{Faculty of Science and Technology, Graduate School of Science and Technology, Hirosaki University, 3, Bunkyo, Hirosaki, Aomori 036-8561, Japan}
\author{K.~Ioka}
\affiliation{Yukawa Institute for Theoretical Physics, Kyoto University, Kitashirakawa Oiwakecho, Sakyo, Kyoto 606-8502, Japan}
\author{W.~Ishizaki}
\affiliation{Institute for Cosmic Ray Research, The University of Tokyo, 5-1-5 Kashiwa-no-Ha, Kashiwa, Chiba 277-8582, Japan}
\author{M.H.~Israel}
\affiliation{Department of Physics and McDonnell Center for the Space Sciences, Washington University, One Brookings Drive, St. Louis, MO 63130-4899, USA}
\author{K.~Kasahara}
\affiliation{Department of Electronic Information Systems, Shibaura Institute of Technology, 307 Fukasaku, Minuma, Saitama 337-8570, Japan}
\author{J.~Kataoka}
\affiliation{Waseda Research Institute for Science and Engineering, Waseda University, 3-4-1 Okubo, Shinjuku, Tokyo 169-8555, Japan}
\author{R.~Kataoka}
\affiliation{National Institute of Polar Research, 10-3, Midori-cho, Tachikawa, Tokyo 190-8518, Japan}
\author{Y.~Katayose}
\affiliation{Faculty of Engineering, Division of Intelligent Systems Engineering, Yokohama National University, 79-5 Tokiwadai, Hodogaya, Yokohama 240-8501, Japan}
\author{C.~Kato}
\affiliation{Faculty of Science, Shinshu University, 3-1-1 Asahi, Matsumoto, Nagano 390-8621, Japan}
\author{N.~Kawanaka}
\affiliation{Hakubi Center, Kyoto University, Yoshida Honmachi, Sakyo-ku, Kyoto, 606-8501, Japan}
\affiliation{Department of Astronomy, Graduate School of Science, Kyoto University, Kitashirakawa Oiwake-cho, Sakyo-ku, Kyoto, 606-8502, Japan}
\author{Y.~Kawakubo}
\affiliation{Department of Physics and Astronomy, Louisiana State University, 202 Nicholson Hall, Baton Rouge, LA 70803, USA}
\author{K.~Kobayashi}
\affiliation{Waseda Research Institute for Science and Engineering, Waseda University, 17 Kikuicho,  Shinjuku, Tokyo 162-0004, Japan}
\affiliation{JEM Utilization Center, Human Spaceflight Technology Directorate, Japan Aerospace Exploration Agency, 2-1-1 Sengen, Tsukuba, Ibaraki 305-8505, Japan}
 \author{K.~Kohri} 
\affiliation{Institute of Particle and Nuclear Studies, High Energy Accelerator Research Organization, 1-1 Oho, Tsukuba, Ibaraki, 305-0801, Japan} 
\author{H.S.~Krawczynski}
\affiliation{Department of Physics and McDonnell Center for the Space Sciences, Washington University, One Brookings Drive, St. Louis, MO 63130-4899, USA}
\author{J.F.~Krizmanic}
\affiliation{Center for Space Sciences and Technology, University of Maryland, Baltimore County, 1000 Hilltop Circle, Baltimore, Maryland 21250, USA}
\affiliation{Astroparticle Physics Laboratory, NASA/GSFC, Greenbelt, Maryland 20771, USA}
\affiliation{Center for Research and Exploration in Space Sciences and Technology, NASA/GSFC, Greenbelt, Maryland 20771, USA}
\author{J.~Link}
\affiliation{Center for Space Sciences and Technology, University of Maryland, Baltimore County, 1000 Hilltop Circle, Baltimore, Maryland 21250, USA}
\affiliation{Astroparticle Physics Laboratory, NASA/GSFC, Greenbelt, Maryland 20771, USA}
\affiliation{Center for Research and Exploration in Space Sciences and Technology, NASA/GSFC, Greenbelt, Maryland 20771, USA}
\author{P.~Maestro}
\email[]{maestro@unisi.it}
\affiliation{Department of Physical Sciences, Earth and Environment, University of Siena, via Roma 56, 53100 Siena, Italy}
\affiliation{INFN Sezione di Pisa, Polo Fibonacci, Largo B. Pontecorvo, 3 - 56127 Pisa, Italy}
\author{P.S.~Marrocchesi}
\affiliation{Department of Physical Sciences, Earth and Environment, University of Siena, via Roma 56, 53100 Siena, Italy}
\affiliation{INFN Sezione di Pisa, Polo Fibonacci, Largo B. Pontecorvo, 3 - 56127 Pisa, Italy}
\author{A.M.~Messineo}
\affiliation{University of Pisa, Polo Fibonacci, Largo B. Pontecorvo, 3 - 56127 Pisa, Italy}
\affiliation{INFN Sezione di Pisa, Polo Fibonacci, Largo B. Pontecorvo, 3 - 56127 Pisa, Italy}
\author{J.W.~Mitchell}
\affiliation{Astroparticle Physics Laboratory, NASA/GSFC, Greenbelt, MD 20771, USA}
\author{S.~Miyake}
\affiliation{Department of Electrical and Electronic Systems Engineering, National Institute of Technology, Ibaraki College, 866 Nakane, Hitachinaka, Ibaraki 312-8508 Japan}
\author{A.A.~Moiseev}
\affiliation{Department of Astronomy, University of Maryland, College Park, Maryland 20742, USA}
\affiliation{Astroparticle Physics Laboratory, NASA/GSFC, Greenbelt, Maryland 20771, USA}
\affiliation{Center for Research and Exploration in Space Sciences and Technology, NASA/GSFC, Greenbelt, Maryland 20771, USA}
\author{M.~Mori}
\affiliation{Department of Physical Sciences, College of Science and Engineering, Ritsumeikan University, Shiga 525-8577, Japan}
\author{N.~Mori}
\affiliation{INFN Sezione di Florence, Via Sansone, 1 - 50019 Sesto, Fiorentino, Italy}
\author{H.M.~Motz}
\affiliation{Faculty of Science and Engineering, Global Center for Science and Engineering, Waseda University, 3-4-1 Okubo, Shinjuku, Tokyo 169-8555, Japan}
\author{K.~Munakata}
\affiliation{Faculty of Science, Shinshu University, 3-1-1 Asahi, Matsumoto, Nagano 390-8621, Japan}
\author{S.~Nakahira}
\affiliation{Institute of Space and Astronautical Science, Japan Aerospace Exploration Agency, 3-1-1 Yoshinodai, Chuo, Sagamihara, Kanagawa 252-5210, Japan}
\author{J.~Nishimura}
\affiliation{Institute of Space and Astronautical Science, Japan Aerospace Exploration Agency, 3-1-1 Yoshinodai, Chuo, Sagamihara, Kanagawa 252-5210, Japan}
\author{G.A.~de~Nolfo}
\affiliation{Heliospheric Physics Laboratory, NASA/GSFC, Greenbelt, MD 20771, USA}
\author{S.~Okuno}
\affiliation{Kanagawa University, 3-27-1 Rokkakubashi, Kanagawa, Yokohama, Kanagawa 221-8686, Japan}
\author{J.F.~Ormes}
\affiliation{Department of Physics and Astronomy, University of Denver, Physics Building, Room 211, 2112 East Wesley Ave., Denver, CO 80208-6900, USA}
\author{N.~Ospina}
\affiliation{Department of Physics and Astronomy, University of Padova, Via Marzolo, 8, 35131 Padova, Italy}\affiliation{INFN Sezione di Padova, Via Marzolo, 8, 35131 Padova, Italy} 
\author{S.~Ozawa}
\affiliation{Quantum ICT Advanced Development Center, National Institute of Information and Communications Technology, 4-2-1 Nukui-Kitamachi, Koganei, Tokyo 184-8795, Japan}
\author{L.~Pacini}
\affiliation{Department of Physics, University of Florence, Via Sansone, 1 - 50019 Sesto, Fiorentino, Italy}
\affiliation{Institute of Applied Physics (IFAC),  National Research Council (CNR), Via Madonna del Piano, 10, 50019 Sesto, Fiorentino, Italy}
\affiliation{INFN Sezione di Florence, Via Sansone, 1 - 50019 Sesto, Fiorentino, Italy}
\author{F.~Palma}
\affiliation{INFN Sezione di Rome ``Tor Vergata'', Via della Ricerca Scientifica 1, 00133 Rome, Italy}
\author{P.~Papini}
\affiliation{INFN Sezione di Florence, Via Sansone, 1 - 50019 Sesto, Fiorentino, Italy}
\author{B.F.~Rauch}
\affiliation{Department of Physics and McDonnell Center for the Space Sciences, Washington University, One Brookings Drive, St. Louis, MO 63130-4899, USA}
\author{S.B.~Ricciarini}
\affiliation{Institute of Applied Physics (IFAC),  National Research Council (CNR), Via Madonna del Piano, 10, 50019 Sesto, Fiorentino, Italy}
\affiliation{INFN Sezione di Florence, Via Sansone, 1 - 50019 Sesto, Fiorentino, Italy}
\author{K.~Sakai}
\affiliation{Center for Space Sciences and Technology, University of Maryland, Baltimore County, 1000 Hilltop Circle, Baltimore, Maryland 21250, USA}
\affiliation{Astroparticle Physics Laboratory, NASA/GSFC, Greenbelt, Maryland 20771, USA}
\affiliation{Center for Research and Exploration in Space Sciences and Technology, NASA/GSFC, Greenbelt, Maryland 20771, USA}
\author{T.~Sakamoto}
\affiliation{College of Science and Engineering, Department of Physics and Mathematics, Aoyama Gakuin University,  5-10-1 Fuchinobe, Chuo, Sagamihara, Kanagawa 252-5258, Japan}
\author{M.~Sasaki}
\affiliation{Department of Astronomy, University of Maryland, College Park, Maryland 20742, USA}
\affiliation{Astroparticle Physics Laboratory, NASA/GSFC, Greenbelt, Maryland 20771, USA}
\affiliation{Center for Research and Exploration in Space Sciences and Technology, NASA/GSFC, Greenbelt, Maryland 20771, USA}
\author{Y.~Shimizu}
\affiliation{Kanagawa University, 3-27-1 Rokkakubashi, Kanagawa, Yokohama, Kanagawa 221-8686, Japan}
\author{A.~Shiomi}
\affiliation{College of Industrial Technology, Nihon University, 1-2-1 Izumi, Narashino, Chiba 275-8575, Japan}
\author{R.~Sparvoli}
\affiliation{University of Rome ``Tor Vergata'', Via della Ricerca Scientifica 1, 00133 Rome, Italy}
\affiliation{INFN Sezione di Rome ``Tor Vergata'', Via della Ricerca Scientifica 1, 00133 Rome, Italy}
\author{P.~Spillantini}
\affiliation{Department of Physics, University of Florence, Via Sansone, 1 - 50019 Sesto, Fiorentino, Italy}
\author{F.~Stolzi}
\affiliation{Department of Physical Sciences, Earth and Environment, University of Siena, via Roma 56, 53100 Siena, Italy}
\affiliation{INFN Sezione di Pisa, Polo Fibonacci, Largo B. Pontecorvo, 3 - 56127 Pisa, Italy}
\author{S.~Sugita}
\affiliation{College of Science and Engineering, Department of Physics and Mathematics, Aoyama Gakuin University,  5-10-1 Fuchinobe, Chuo, Sagamihara, Kanagawa 252-5258, Japan}
\author{J.E.~Suh} 
\affiliation{Department of Physical Sciences, Earth and Environment, University of Siena, via Roma 56, 53100 Siena, Italy}
\affiliation{INFN Sezione di Pisa, Polo Fibonacci, Largo B. Pontecorvo, 3 - 56127 Pisa, Italy}
\author{A.~Sulaj} 
\affiliation{Department of Physical Sciences, Earth and Environment, University of Siena, via Roma 56, 53100 Siena, Italy}
\affiliation{INFN Sezione di Pisa, Polo Fibonacci, Largo B. Pontecorvo, 3 - 56127 Pisa, Italy}
\author{M.~Takita}
\affiliation{Institute for Cosmic Ray Research, The University of Tokyo, 5-1-5 Kashiwa-no-Ha, Kashiwa, Chiba 277-8582, Japan}
\author{T.~Tamura}
\affiliation{Kanagawa University, 3-27-1 Rokkakubashi, Kanagawa, Yokohama, Kanagawa 221-8686, Japan}
\author{T.~Terasawa}
\affiliation{RIKEN, 2-1 Hirosawa, Wako, Saitama 351-0198, Japan}
\author{S.~Torii}
\affiliation{Waseda Research Institute for Science and Engineering, Waseda University, 17 Kikuicho,  Shinjuku, Tokyo 162-0004, Japan}
\author{Y.~Tsunesada}
\affiliation{Division of Mathematics and Physics, Graduate School of Science, Osaka City University, 3-3-138 Sugimoto, Sumiyoshi, Osaka 558-8585, Japan}
\author{Y.~Uchihori}
\affiliation{National Institutes for Quantum and Radiation Science and Technology, 4-9-1 Anagawa, Inage, Chiba 263-8555, JAPAN}
\author{E.~Vannuccini}
\affiliation{INFN Sezione di Florence, Via Sansone, 1 - 50019 Sesto, Fiorentino, Italy}
\author{J.P.~Wefel}
\affiliation{Department of Physics and Astronomy, Louisiana State University, 202 Nicholson Hall, Baton Rouge, LA 70803, USA}
\author{K.~Yamaoka}
\affiliation{Nagoya University, Furo, Chikusa, Nagoya 464-8601, Japan}
\author{S.~Yanagita}
\affiliation{College of Science, Ibaraki University, 2-1-1 Bunkyo, Mito, Ibaraki 310-8512, Japan}
\author{A.~Yoshida}
\affiliation{College of Science and Engineering, Department of Physics and Mathematics, Aoyama Gakuin University,  5-10-1 Fuchinobe, Chuo, Sagamihara, Kanagawa 252-5258, Japan}
\author{K.~Yoshida}
\affiliation{Department of Electronic Information Systems, Shibaura Institute of Technology, 307 Fukasaku, Minuma, Saitama 337-8570, Japan}

\collaboration{CALET Collaboration}

\date{\today}

\begin{abstract}
In this paper, we present the measurement of the energy spectra of carbon and oxygen in cosmic rays 
based on observations with the Calorimetric Electron Telescope (CALET) on the International Space Station from October 2015 to October 2019.
Analysis, including the detailed assessment of systematic uncertainties, and results are reported. 
The energy spectra are measured in kinetic energy per nucleon from 10 GeV$/n$ to 2.2 TeV$/n$
with an all-calorimetric instrument with a total thickness corresponding to 1.3 nuclear interaction length. 
The observed carbon and oxygen fluxes 
show a spectral index change of $\sim$0.15 around 200 GeV$/n$
established with a significance $>3\sigma$. 
They have  the same energy dependence with a constant C/O flux ratio $0.911\pm 0.006$  above 25 GeV$/n$.
The spectral hardening is consistent with that measured by AMS-02, but the absolute normalization of the flux
is about 27\% lower, though in agreement with observations from previous experiments including  the PAMELA spectrometer 
\textcolor{black}{and the calorimetric balloon-borne experiment CREAM.}
\end{abstract}

\maketitle

\section{Introduction}
Direct measurements  
of charged cosmic rays (CR) 
provide information on their origin, acceleration and propagation in the Galaxy.
Search for possible charge-dependent cutoffs in the nuclei spectra, hypothesized to explain
the “knee” in the all-particle spectrum ~\cite{Lagage,Berezhko,Horandel,Parizot}, 
can be pursued at very large energy by magnetic spectrometers with sufficient rigidity coverage (MDR) or by calorimetric instruments 
equipped with charge detectors capable of single element resolution.
Furthermore, recent observations indicating a spectral hardening in proton and He spectra ~\cite{AMS-PROTON, AMS-HE, PAMELA-PHE, CREAM1, CREAM3,DAMPE-PROTON} 
as well as in heavy nuclei spectra ~\cite{AMS-CO, CREAM2HARD, TRACER2009, TRACER2011} around a few hundred GeV$/n$,
compelled a revision of the standard paradigm of galactic CR based on diffusive shock acceleration 
in supernova remnants followed by propagation in galactic magnetic fields,
and prompted an intense theoretical activity to interpret these unexpected spectral features
\cite{Serpico,Malkov,Bernard,Tomassetti,Drury,Blasi2012,Evoli,Ohira2011,Ohira2016,Vladimirov,Ptuskin,Thoudam}.
The CALorimetric Electron Telescope (CALET) ~\cite{CALET, CALET2, CALET3} is a  space-based instrument optimized for the measurement
of the all-electron spectrum ~\cite{CALET-ELE2017,CALET-ELE2018}, which can also measure 
individual chemical elements in CR from proton to iron and above in the energy range up to $\sim$1 PeV. 
CALET recently confirmed the spectral hardening in the proton spectrum by accurately measuring 
its power-law spectral index over the wide energy range from  50 GeV to 10 TeV ~\cite{CALET-PROTON}.\\
In this Letter, we present a new direct measurement of the CR carbon and oxygen spectra from 10 GeV$/n$ to 2.2 TeV$/n$,
based on the data collected  by CALET from October 13, 2015 to October 31, 2019 aboard the International Space Station (ISS).
\section{CALET Instrument}
CALET consists of a CHarge Detector (CHD),
a finely segmented pre-shower IMaging Calorimeter (IMC), and a Total AbSorption Calorimeter (TASC).
CHD is comprised of two hodoscopes made of 14 plastic scintillator paddles each, arranged in orthogonal layers (CHDX, CHDY).  
The CHD can resolve individual chemical elements from $Z=1$ to $Z=40$, with excellent charge resolution.
The IMC consists of 7 tungsten plates inserted between eight double layers of
 1 mm$^2$ cross-section scintillating fibers, arranged in belts along orthogonal directions 
and individually read out by multianode photomultiplier tubes.
Its fine granularity and imaging capability allow 
an accurate particle tracking and \textcolor{black}{an independent charge measurement via multiple 
samples of the particle ionization energy loss ($dE/dx$)  in each fiber}.
The TASC is a homogeneous calorimeter made of lead-tungstate (PbWO$_4$)  bars
arranged in 12 layers.  
The crystal bars in the top layers are \textcolor{black}{read out by photomultiplier tubes}, while
a dual photodiode/avalanche-photodiode (PD/APD) system is used for each channel in the remaining layers.
A dynamic range of more than six \textcolor{black}{orders} of magnitude is covered using  a front-end electronics with dual gain range for each photosensor. 
The total thickness of the instrument is equivalent to 30 radiation \textcolor{black}{lengths} and 1.3 nuclear interaction \textcolor{black}{lengths}. 
A more complete description of the instrument can be found in the Supplemental Material (SM) of Ref.~\cite{CALET-ELE2017}.\\
CALET was launched on August 19, 2015  
and  installed on the Japanese Experiment Module Exposure Facility  of the ISS. 
The on-orbit commissioning phase aboard the ISS was successfully completed in the first days of October 2015, 
and since then the instrument has been taking science data continuously~\cite{CALET2018}. 
\section{Data analysis}
We have analyzed flight data (FD) collected in 1480 days of CALET operation.
The total observation live time for  the high-energy (HE) shower trigger is 
\textcolor{black}{
$T=3.00\times10^4$} hours, corresponding to 84.5\% of total observation time. \\
Raw data are corrected for non-uniformity in light output, time and temperature dependence, gain differences among the channels. The latter are individually calibrated on orbit by
using penetrating proton and He particles, selected by a dedicated trigger mode ~\cite{CALET2017,niita}.
After 
calibrations, each CR particle track is reconstructed and a charge and an energy are assigned for each event.

Monte Carlo (MC) simulations,  reproducing the detailed detector configuration, physics processes, as well as detector signals,
are based on the EPICS simulation package~\cite{EPICS, EPICSurl} and employ 
the hadronic interaction model DPMJET-III~\cite{dpmjet3prl}.
An independent analysis based on FLUKA~\cite{FLUKA,FLUKA2} is also performed to assess the systematic uncertainties.

The CR particle direction and its entrance point in the instrument are reconstructed by a  track finding and fitting algorithm 
based on a combinatorial Kalman filter \cite{paolo2017}, which is able to identify 
the incident  track  in the presence of a background of secondary tracks backscattered from TASC.
The angular resolution is $\sim0.1^\circ$  for C and O nuclei and the spatial resolution 
on the determination of the impact point on CHD is $\sim$220 $\mu$m.

The identification of the particle charge $Z$ is based  on the measurements of the ionization deposits in CHD and IMC.
The particle trajectory is used to identify the CHD paddles and IMC fibers traversed by the primary particle
and to determine the path length correction to be applied to the signals to extract the $dE/dx$ samples. 
Three independent $dE/dx$ measurements are obtained, one for each CHD layer and the third by averaging
the samples (at most eight) along the track in the top half of IMC. 
Calibration curves of $dE/dx$ are built by fitting FD subsets for each nuclear species to a function
of $Z^2$ by using a "halo" model \cite{GSI}. 
These curves are then used 
to reconstruct   three charge values   ($Z_{\rm CHDX}$, $Z_{\rm CHDY}$, $Z_{\rm IMC}$) from the measured $dE/dx$  on an event-by-event basis \cite{CALET-CO}.
For high-energy showers, the charge peaks are corrected for the systematical shift to higher values (up to 0.15 $e$) with respect to the nominal charge positions, 
due to the large amount  of shower particle tracks backscattered from TASC whose signals add up to the primary particle ionization signal.
A charge distribution obtained by averaging $Z_{\rm CHDX}$ and $Z_{\rm CHDY}$  is shown 
in Fig.~S1 of the SM \cite{PRL-SM}.
The charge resolution $\sigma_Z$ is $\sim0.15\, e$ (charge unit) for CHD and $\sim0.24\, e$
for IMC, respectively, in the elemental range from B to O.

The shower energy $E_{\rm TASC}$ of each event is calculated as
the sum of the energy deposits of all the TASC channels, after stitching the adjacent gain ranges of each PD/APD.
\textcolor{black}{
The energy response of TASC was studied in a beam test carried out at CERN-SPS in 2015
with accelerated ion fragments  of 13, 19 and 150 GeV/c momentum per nucleon ~\cite{akaike2015}. The MC simulations were tuned using the beam test results
as described in the {\em Energy measurement} section of the SM \cite{PRL-SM}.
}

Carbon and oxygen candidates are identified among events selected by the onboard HE shower trigger,  
based on the coincidence of the summed signals of the last two IMC layers in each view and the top TASC layer (TASCX1). 
Consistency between MC and FD for triggered events is obtained by an offline trigger with higher thresholds 
(50 and 100  times a minimum ionizing particle (MIP) signal for IMC and TASC, respectively)
than the onboard trigger removing possible effects due to residual non-uniformity of the detector gain.\\   
In order to reject possible events triggered by particles entering the TASC from
lateral sides or with significant lateral leakage, 
the  energy deposits in the first TASC layer (TASCX1) and in all the lateral  bars 
are required to be less than 40\% of  $E_{\rm TASC}$.
Late-interacting events in the bottom half of TASC are rejected by requiring that 
 the energy deposit in the last layer is $<0.4\times E_{\rm TASC}$,
and the layer, where the longitudinal shower development reaches  20\% of $E_{\rm TASC}$, occurs in the upper half of TASC. \\
Events with one well-fitted track crossing the whole detector from CHD top to the TASC bottom layer 
and at least 2 cm away from the edges in TASCX1
are then selected.
The fiducial geometrical factor  for this category of events is $S\Omega \sim$510 cm$^2$sr, corresponding to about 50\% of the total CALET acceptance.\\
Carbon and oxygen candidates are selected by applying  window cuts, centered on the nominal charge values ($Z=6, 8$), 
of half-width $0.4\, e$  for $Z_{CHDX}$ and $Z_{CHDY}$, and  $2 \sigma_Z$ for $Z_{IMC}$, respectively.
Particles undergoing a charge-changing nuclear interaction in the upper part of the instrument (Fig.~S2 of the SM \cite{PRL-SM})
are removed by the three combined charge selections and by requiring
the consistency, within 30\%, between the mean values of $dE/dx$ measurements in the first four layers in each IMC view.\\
Distributions of $E_{\rm TASC}$ for C and O selected candidates  are shown in Fig.~S3 of SM \cite{PRL-SM}, corresponding to 
6.154$\times 10^5$ C and  1.047$\times 10^6$ O events,  respectively. 
In order to take into account the relatively limited energy resolution 
\textcolor{black}{(Fig.~S4 of the SM \cite{PRL-SM})}
energy unfolding is necessary to correct for bin-to-bin migration effects. 
In this analysis, we used the Bayesian approach \cite{Ago} 
implemented in the RooUnfold package \cite{ROOUNFOLD} in ROOT \cite{ROOT}. 
Each element of the response matrix represents
the probability that primary nuclei in a certain energy interval of the CR spectrum produce 
an energy deposit in a given bin of $E_{\rm TASC}$.
The response matrix \textcolor{black}{(Fig.~S5 of the SM \cite{PRL-SM})}
is derived using MC simulation after applying the same selection procedure as for FD. \\
The energy spectrum is obtained from the unfolded energy distribution as follows:
\begin{equation}
\Phi(E) = \frac{N(E)}{\Delta E\;  \varepsilon(E) \;  S\Omega \;  T }
\label{eq_flux}
\end{equation}
\begin{equation}
N(E) = U \left[N_{obs}(E_{\rm TASC}) - N_{bg}(E_{\rm TASC}) \right]
\end{equation}
where $\Delta E$ denotes energy bin width, 
 $E$ the particle kinetic energy, calculated as the geometric mean of the lower and upper bounds of the bin, 
$N(E)$ is the bin content in the unfolded distribution,
$\varepsilon (E)$ the total selection efficiency (Fig.~S6 of the SM \cite{PRL-SM}), 
$U()$  the unfolding procedure, 
$N_{obs}(E_{\rm TASC})$ the bin content of observed energy distribution (including background),
$N_{bg}(E_{\rm TASC})$ the bin content of background events in the observed energy distribution.
Background contamination from different nuclear species misidentified as C or O is shown in Fig.~S3 of the SM \cite{PRL-SM}.  
A contamination fraction $N_{bg}/N_{obs} <0.1\%$ is found in all energy bins with $E_{\rm TASC}<10^3$ GeV, and between 0.1\% and 1\% 
for $E_{TASC} > 10^3$ GeV.
\section{Systematic Uncertainties}
In this analysis, dominant sources of systematic uncertainties  include
trigger efficiency, energy response, event selection, unfolding procedure, MC model.\\
HE trigger efficiency  as a function of $E_{\rm TASC}$ was inferred from the data taken with a minimum bias trigger. 
HE efficiency curves for C and O  are consistent with predictions from MC simulations, as shown in Fig.~S7 of the SM \cite{PRL-SM}. 
In order to study the flux stability against offline trigger efficiency, the threshold applied to TASCX1 signal was scanned between 100 and 150 MIP signal. 
The corresponding systematic errors range between -4.2\% (-3.1\%) and 3.7\% (7.3\%) for C (O) depending on the energy bin.\\
%
The systematic error related to charge identification was studied by varying 
the width of  the window cuts between 0.35\,$e$ and 0.45\,$e$  for CHD and between 1.75\,$\sigma_Z$ and 2.2\,$\sigma_Z$ for IMC.
That results in a flux variation depending on the energy bin, which is less than 1\% below 250 $\text{GeV}/n$ and  few percent above. 
\textcolor{black}{The ratio of events selected by IMC charge cut to the ones selected with CHD  in different $E_{\rm TASC}$  intervals  turned out to be consistent in FD and MC.}
\\
Possible inaccuracy of track reconstruction could affect  
the determination of the geometrical acceptance. The contamination due to 
off-acceptance events which are mis-reconstructed in the fiducial acceptance
was estimated with MC to be $\sim 1\%$ at 10 GeV$/n$ and decrease to less than $0.1\%$ above 60 GeV$/n$. 
To investigate the uncertainty in the definition of the acceptance, restricted  acceptance (up to 20\% of nominal one)
regions were also studied. The corresponding fluxes are consistent within statistical fluctuations.\\
A different tracking procedure, described in Ref.~\cite{akaike2019}, was also used
to study possible systematic uncertainties in tracking efficiency. 
Results are consistent with those obtained with the Kalman filter algorithm, hence we 
 consider negligible this source of systematic error.\\
The uncertainty in the energy scale  
 is $\pm2$\% and depends on the accuracy of the beam test calibration. 
It causes a rigid shift of the measured energies, affecting 
the absolute normalization of the C and O spectra by $^{+2.6\%}_{-2.8\%}$, but not their shape.
As the beam test model was not identical to the instrument now in orbit, the difference in the 
spectrum obtained with either configuration was modeled and included in the systematic error.\\
Other energy-independent systematic uncertainties affecting the normalization
include live time (3.4\%, as explained in the SM of Ref.~\cite{CALET-ELE2017})
and long-term stability of the charge measurements ($<0.4\%$). \\
The uncertainties due to the unfolding procedure 
were evaluated by using  different response matrices, computed by varying the spectral index (between -2.9 and -2.5) of the generation spectrum of MC simulations,
and the  Singular Value Deconvolution method, instead of the Bayesian approach, in RooUnfold software \cite{ROOUNFOLD}.\\
Since it is not possible to validate MC simulations with  beam test data in the high-energy region,
a comparison between different MC models,  i.e. EPICS and FLUKA,  was performed. 
We found that the total selection efficiencies for C and O determined with the two models are in agreement within $<1.5\%$  over the whole energy range, 
but the energy response matrices differ significantly in the low and high energy regions.
The resulting fluxes show maximum discrepancies of 9\% (7.8\%) 
and 9.2\% (12.2\%), respectively, in the first and last energy bin for C (O), 
while they are consistent within 6.6\% (6.2\%) elsewhere.
This is the dominant source of systematic uncertainties.\\
Materials traversed by nuclei in IMC are mainly composed of carbon, aluminum and tungsten. 
Possible uncertainties in the inelastic cross sections in simulations  or discrepancies in the material description 
 might affect the flux normalization.
We have checked that hadronic interactions are well simulated in the detector, by measuring the 
survival probabilities  of C and O nuclei at different depths in IMC, as described in the SM \cite{PRL-SM}.
The survival probabilities are in agreement with MC prediction within $<1\%$ (Fig.~S8 of the SM  \cite{PRL-SM}).\\
Background contamination from different nuclear species estimated with FLUKA and EPICS simulations differ by less than $1\%$.\\
The energy dependence of all the systematic uncertainties for C and O  is shown in  Fig.~S9 of the SM~\cite{PRL-SM}. 
The total systematic error is computed as the sum in quadrature of all the sources of systematics in each energy bin. 
\section{Results}
The energy spectra of carbon and oxygen and their flux ratio measured with CALET in an energy range from 10 GeV$/n$ to 2.2 TeV$/n$ are shown in Fig.~\ref{fig:flux}, 
where current uncertainties that include statistical and systematic errors are bounded within a gray band. 
CALET spectra are compared with results from space-based  \cite{HEAO,CRN,PAMELA-C,AMS-CO, NUCLEON} 
and balloon-borne \cite{Simon,ATIC2,TRACER, CREAM2} experiments.
The measured C and O fluxes and flux ratio with statistical and systematic errors are tabulated in Tables I, II and III 
of the SM \cite{PRL-SM}.
\begin{figure} \centering
\includegraphics[width=\hsize]{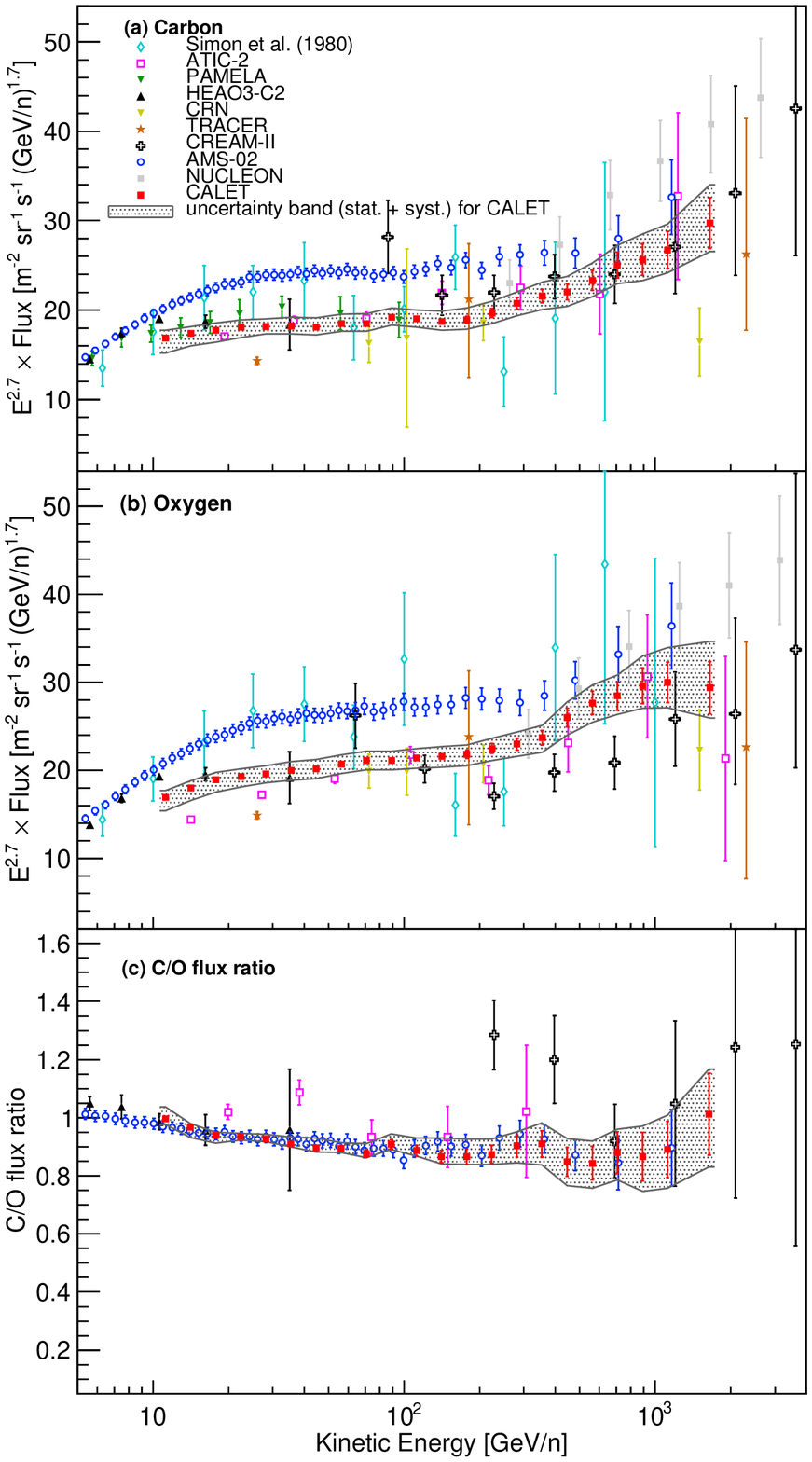} 
\caption{\scriptsize CALET (a) carbon and (b) oxygen flux (multiplied by $E^{2.7}$) and (c)  ratio of carbon to oxygen fluxes, as a function of kinetic energy $E$.
Error bars of CALET data (red) represent the statistical uncertainty only, while the  gray band indicates the quadratic sum of statistical and systematic errors. 
Also plotted are other direct measurements \cite{HEAO, CRN, ATIC2, TRACER, AMS-CO, PAMELA-C, CREAM2, NUCLEON, Simon}.
\textcolor{black}{An enlarged version of the figure is available as Fig.~S10 in the SM \cite{PRL-SM}}. }. 
\label{fig:flux}
\end{figure}\noindent
Our spectra are consistent with PAMELA ~\cite{PAMELA-C} and most previous experiments \cite{HEAO, CRN, ATIC2, TRACER, CREAM2}, 
but the absolute normalization is in tension with AMS-02 \cite{AMS-CO}. 
However we notice that C/O ratio (Fig.~\ref{fig:flux} (c)) is consistent with the one measured by AMS-02. 
In Fig.~S11 
of the SM \cite{PRL-SM}, it is shown that CALET and AMS-02 C and O spectra have very similar shapes
but they differ in the absolute normalization, which is lower for CALET by about 27\% for both C and O. \\
Figure \ref{fig:COfit} shows the fits to CALET carbon and oxygen data with a double power-law function (DPL, Eq.~S1  in SM \cite{PRL-SM}) 
above 25 GeV$/n$. A single power-law function (SPL, Eq.~S2 in SM \cite{PRL-SM}))  fitted to data 
in the energy range [25, 200] GeV$/n$ and extrapolated above 200 GeV$/n$ is also shown for comparison.
The effect of systematic uncertainties in the measurement of the energy spectrum is modeled 
 in the $\chi^2$ minimization function with a set of 6 nuisance parameters as explained in detail in the SM \cite{PRL-SM}.
The DPL fit to the C spectrum yields a spectral index $\gamma = -2.663\pm0.014$ at energies below the transition region $E_0 = (215\pm54)$ GeV$/n$
and a  spectral index increase $\Delta\gamma = 0.166\pm0.042$ above, with $\chi^2/$d.o.f. = 9.0/8.
For oxygen, the fit yields $\gamma = -2.637\pm0.009$,  $E_0 = (264\pm53)$ GeV$/n$, $\Delta\gamma = 0.158\pm0.053$, with $\chi^2/$d.o.f. = 3.0/8.
SPL fits to CALET carbon and oxygen data above 25 GeV$/n$ are shown in Figs.~S12 and S13 
of the SM \cite{PRL-SM}.
They give $\gamma = -2.626\pm0.010$ with $\chi^2/$d.o.f. = 27.5/10  for C,  and $\gamma=-2.622\pm0.008$ with $\chi^2/$d.o.f. = 15.9/10 for O, respectively.
A frequentist test statistic $\Delta\chi^2$ is computed from the difference in $\chi^2$ between the fits with SPL and DPL functions.
For carbon (oxygen), $\Delta\chi^2=18.5$ (12.9) with 2 d.o.f. (i.e. the number of additional free parameters in DPL fit with respect to SPL fit)
implies that the significance of the hardening of the C (O) spectrum exceeds the 3$\sigma$ level.
\textcolor{black}{We also checked that the spectral hardening is not an artifact of the energy binning and unfolding, by increasing the 
bin width by a factor 2.5 as shown in Fig.~S14  of the SM \cite{PRL-SM}. 
The resulting flux difference is negligible when compared with our estimated systematic uncertainties.} \\
In order to study the energy dependence of the spectral index in a model independent way, 
the spectral index $\gamma$ is calculated by a fit of $d[\log(\Phi)]/d[\log(E)]$ in energy windows centered in each bin
and including the neighbor $\pm$3 bins. The results in Fig.~\ref{fig:SpectralIndex} show
that carbon and oxygen fluxes harden 
in a similar way above a few hundred GeV$/n$.
The carbon to oxygen flux ratio 
is well fitted to a constant value of $0.911\pm 0.006$  above 25 GeV$/n$ (Fig.~S15  of the SM \cite{PRL-SM}), 
indicating that the two fluxes have the same energy dependence. 
\begin{figure} \centering
\includegraphics[height=9.1cm, width=9.1cm]{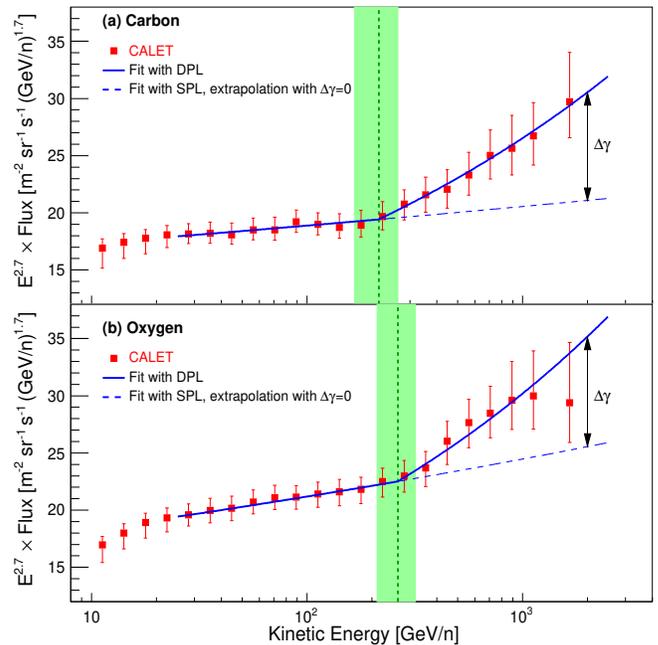}
\caption{\scriptsize  Fit of the CALET (a) C and (b) O energy spectra with a DPL  function (blue line) in the energy range [25, 2000] GeV$/n$. 
The flux is multiplied by $E^{2.7}$ where $E$ is the kinetic energy per nucleon.
Error bars of CALET data points represent the sum in quadrature of statistical and systematic uncertainties. 
The dashed blue lines represent the extrapolation of a SPL function fitted to data in the energy range [25, 200] GeV$/n$.
$\Delta\gamma$ is the change of the spectral index above the transition energy $E_0$, represented by the vertical green dashed line. The error interval for $E_0$ from the DPL fit
is shown by the green band.}
\label{fig:COfit}
\end{figure}\noindent
\begin{figure} 
\centering
\includegraphics[width=9.0cm, height=9.0cm]{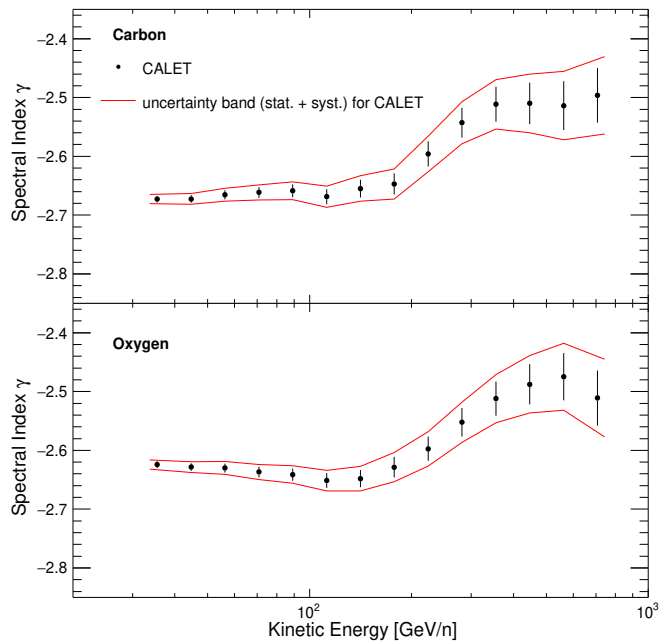}
\caption{\scriptsize  Energy dependence of the spectral index calculated within a sliding energy window for CALET  (a) C and (b) O data. The spectral index is determined for each bin
by fitting the data using $\pm$3 bins. Red curves indicate the uncertainty range including systematic errors. }
\label{fig:SpectralIndex}
\end{figure}\noindent
\vspace{-0.3cm}
\section{Conclusion}
\vspace{-0.3cm}
With a calorimetric apparatus  in low Earth orbit, CALET has measured
the energy spectra of carbon and oxygen nuclei in CR and their flux ratio from 10 GeV$/n$ to 2.2 TeV$/n$.
Our observations  allow to exclude a single power law spectrum for C and O by more than $3\sigma$;  
they show a spectral index increase $\Delta\gamma = 0.166\pm0.042$ for C and $\Delta\gamma = 0.158\pm0.053$ for O above 200 GeV$/n$, 
and the same energy dependence for C and O fluxes with a constant C/O flux ratio $0.911\pm 0.006$  above 25 GeV$/n$.
These results are consistent with the ones reported by AMS-02. 
However the absolute normalization of our data is significantly lower than AMS-02, but in agreement with previous experiments. 
Improved statistics and refinement of the analysis with additional data collected during the lifetime 
of the mission
will allow to extend the measurements at higher energies and improve the spectral analysis,
contributing to a better understanding of the origin of the spectral hardening. 
\section{Acknowledgments}
\vspace{-0.5cm}
\begin{acknowledgments}
We gratefully acknowledge JAXA’s contributions to the development of CALET and to the operations onboard the International Space Station.
We also wish to express our sincere gratitude to Agenzia Spaziale Italiana (ASI) and NASA for their support of the CALET project.
This work was supported in part by JSPS Grant-in-Aid for Scientific Research (S) Number 26220708 and 19H05608, 
JSPS Grant-in-Aid for Scientific Research (B) Number 17H02901, and by the
MEXT-Supported Program for the Strategic Research Foundation at Private Universities (2011-2015)
(No. S1101021) at Waseda University.
The CALET effort in Italy is supported by ASI under agreement 2013-018-R.0 and its amendments.
The CALET effort in the United States is supported by NASA through Grants No. NNX16AB99G, No. NNX16AC02G, and No. NNH14ZDA001N-APRA-0075.
\end{acknowledgments}
\providecommand{\noopsort}[1]{}\providecommand{\singleletter}[1]{#1}%

\widetext
\clearpage
\begin{center}
\end{center}
\setcounter{equation}{0}
\setcounter{figure}{0}
\setcounter{table}{0}
\setcounter{page}{1}
\makeatletter
\renewcommand{\theequation}{S\arabic{equation}}
\renewcommand{\thefigure}{S\arabic{figure}}
\renewcommand{\bibnumfmt}[1]{[S#1]}
\renewcommand{\citenumfont}[1]{S#1}
\begin{center}
\textbf{\large Direct Measurement of the Cosmic-Ray Carbon and Oxygen spectra\\ from 10 GeV/n to 2.2 TeV/n
 with the Calorimetric Electron Telescope \\on the International Space Station \\  
\vspace*{0.2cm}
SUPPLEMENTAL MATERIAL}	\\
\vspace*{0.2cm}
(CALET collaboration) 
\end{center}
\vspace*{1cm}
Supplemental material concerning ``Direct Measurement of the Cosmic-Ray Carbon and Oxygen spectra from 10 GeV$/n$ to 2.2 TeV$/n$  
 with the Calorimetric Electron Telescope on the International Space Station''
\vspace*{1cm}
\clearpage
\begin{figure}[h] \centering
\subfigure[]
{
\includegraphics[width=12cm, height=9cm]{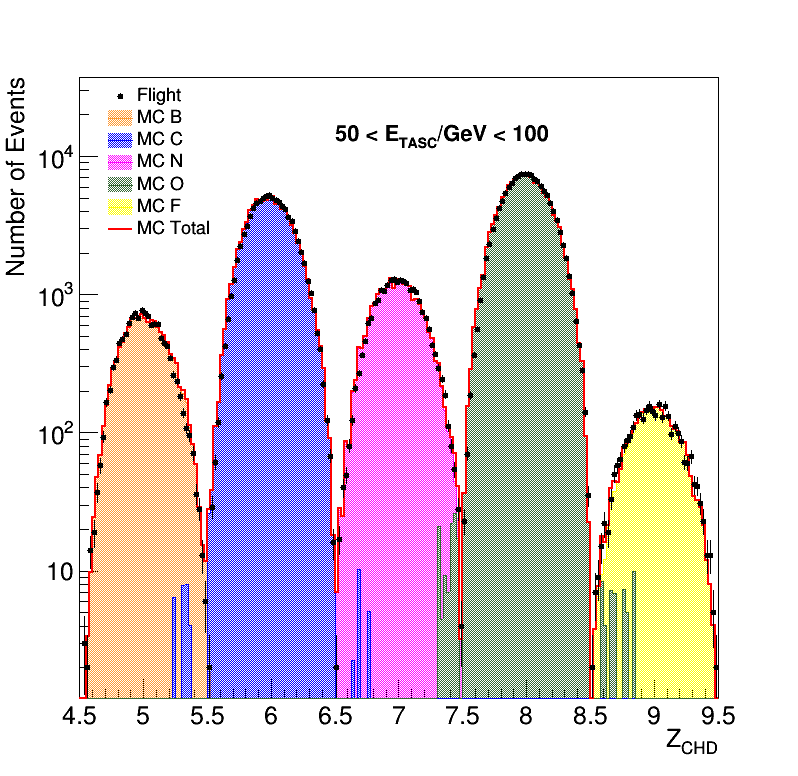}  
  \label{fig:Z_CHD_BCNO2_SMa}
}
\subfigure[]
{
\includegraphics[width=12cm, height=9cm]{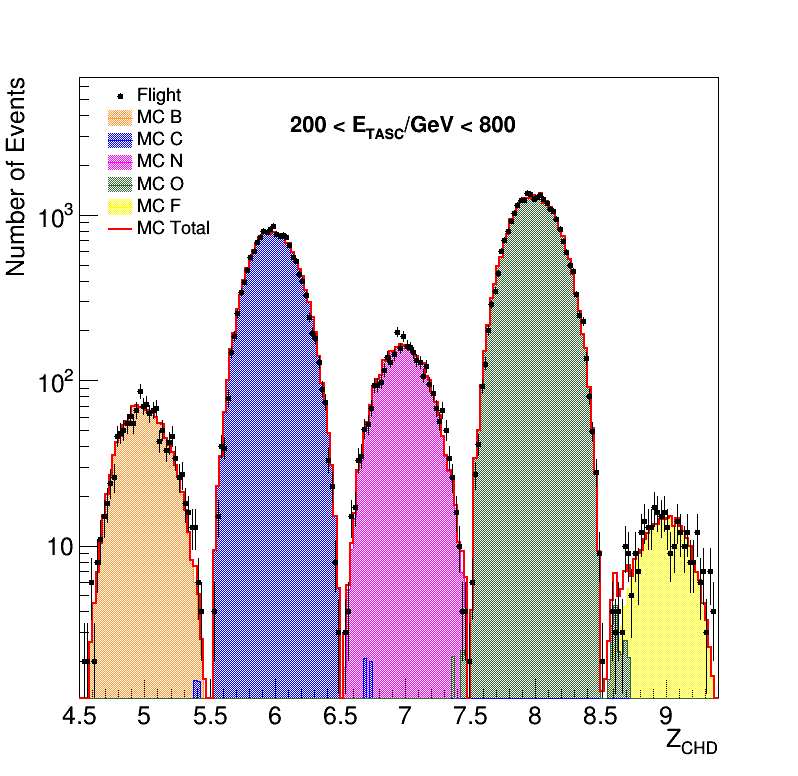}  
  \label{fig:Z_CHD_BCNO2_SMb}
}
\caption{Charge distributions from the combined CHD layers in the elemental region between B and F. 
Events are selected with (a) $50 < E_{\rm TASC} < 100$  GeV and (b) $200 < E_{\rm TASC} < 800$ GeV and a measured charge in IMC consistent with $Z_{\rm CHD}$.
FD (black dots) are compared to MC samples.
\textcolor{black}{The distributions shown in these plots are only representative of the charge resolution while the relative elemental abundances are not meaningful, 
because of the different trigger efficiency for different nuclear species (Fig.~\ref{fig:HEeff}) and different  intervals of
primary energy per nucleon selected.}
}
\label{fig:Z_CHD_BCNO2_SM}
\end{figure}\noindent
\clearpage
\begin{figure} \centering
\includegraphics[width=9cm, height=9cm]{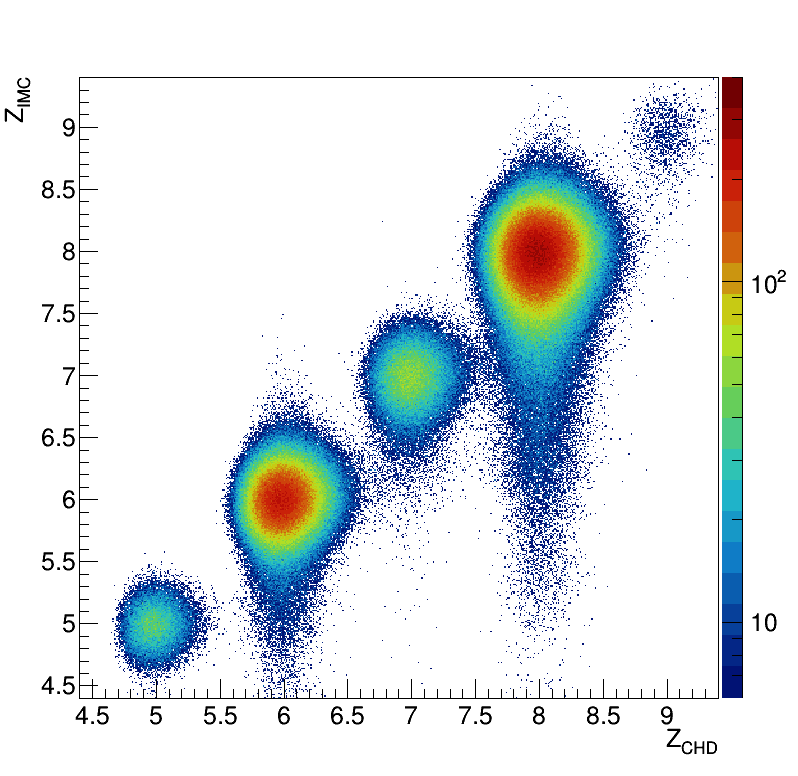}
\caption{Crossplot of IMC versus CHD reconstructed charges in the elemental range between B and F.  
Events with C and O nuclei undergoing a charge-changing nuclear interaction upstream the IMC are clearly visible in the tail of their 
drop-shaped distributions extending to lower $Z_{IMC}$ values. 
These events are removed in the analysis by requiring
the consistency, within 30\%, between the mean values of $dE/dx$ measurements in the first four layers in each IMC view
and then  by applying a window cut to $Z_{\rm IMC}$  of half-width 2$\sigma_{Z}$
centered around the nominal charge values.
}
\label{fig:ZIMCCHD}
\end{figure}\noindent
\begin{figure}
\begin{center}
\subfigure[]
{
\includegraphics[width=8.5cm, height=6cm]{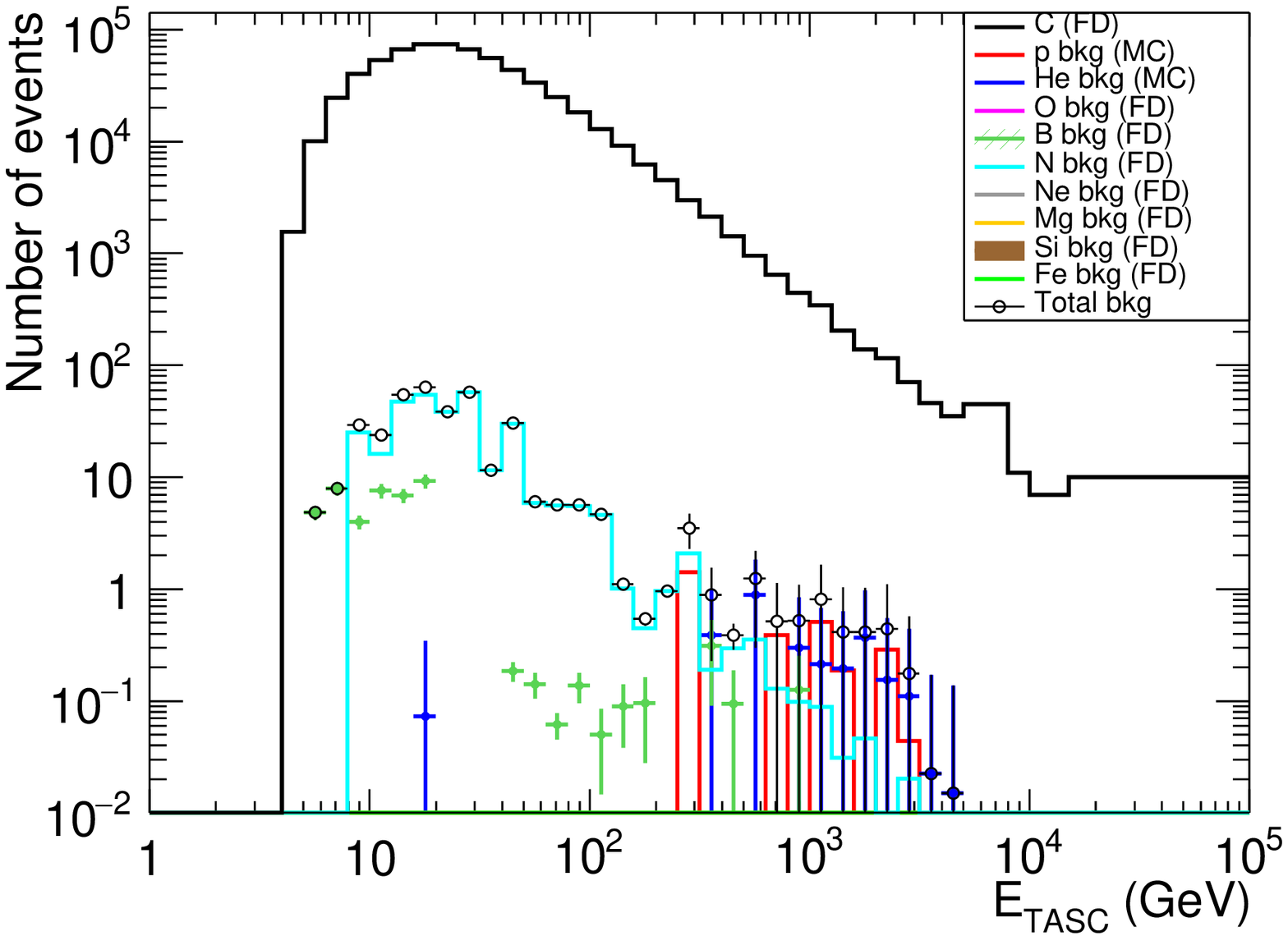}
\label{fig:TASCedepC}                            
}
\subfigure[]
{
\includegraphics[width=8.5cm, height=6cm]{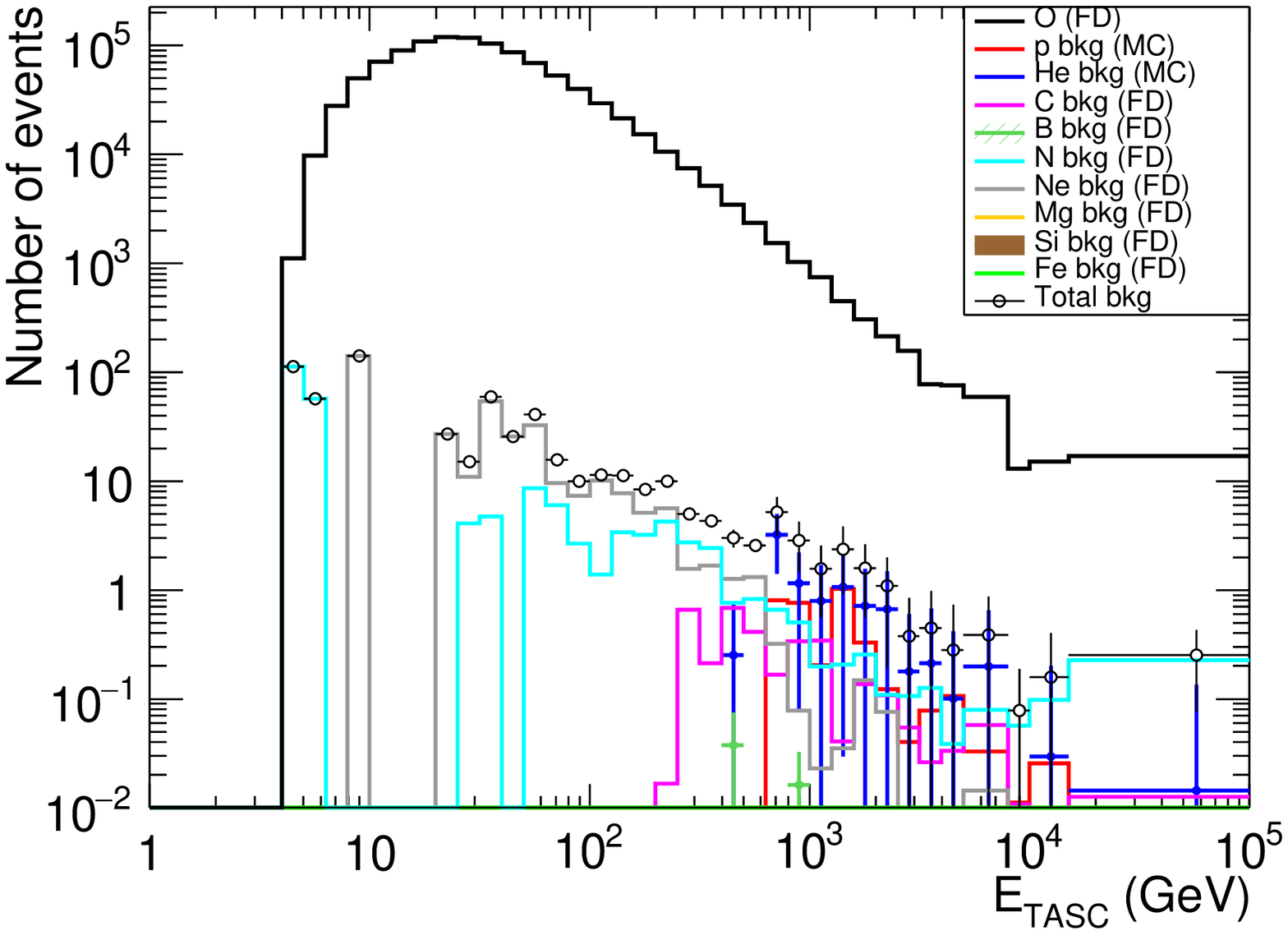}
\label{fig:TASCedepO}                            
}
\subfigure[]
{
\includegraphics[width=8.5cm, height=6cm]{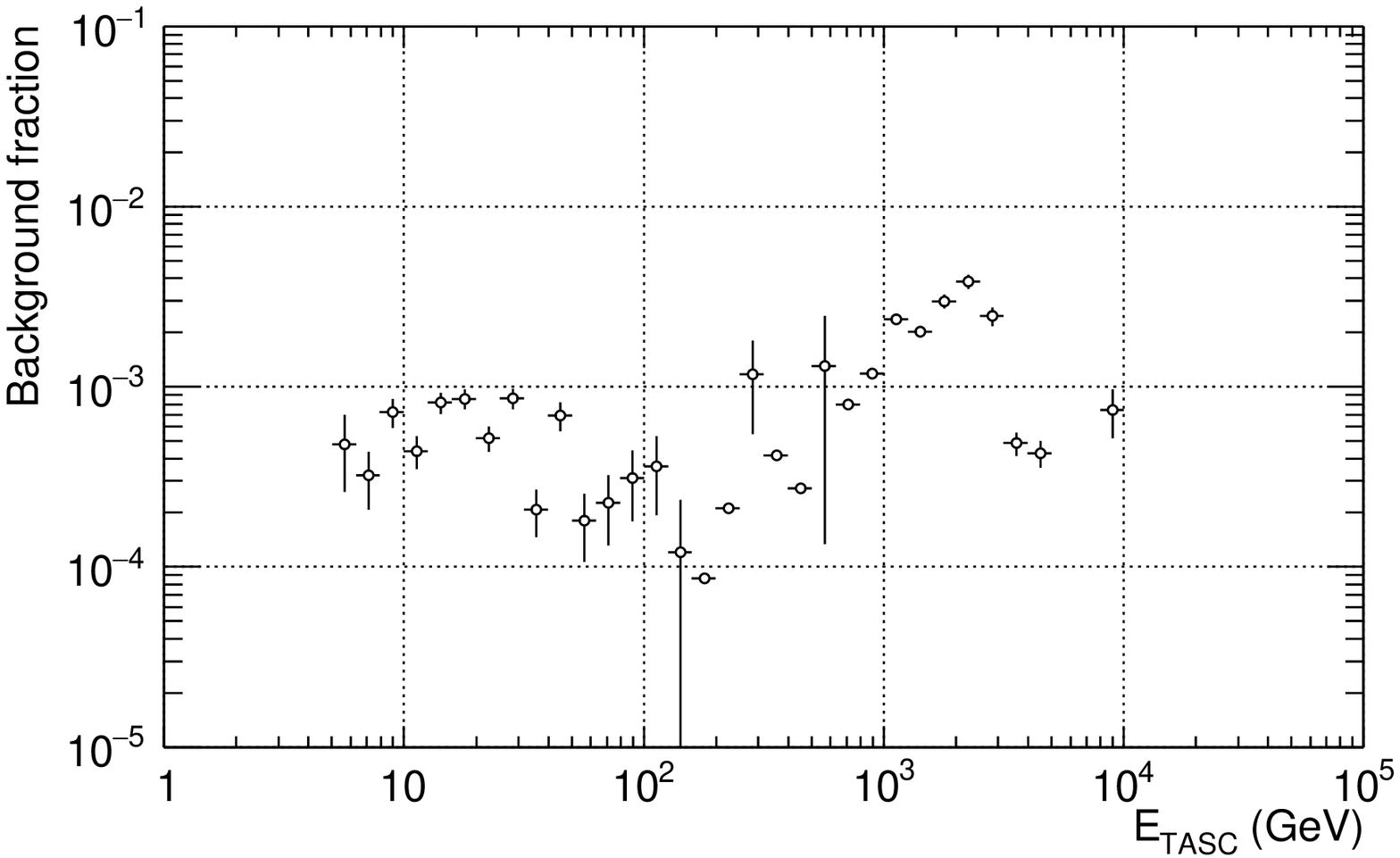}
\label{fig:bkgfracC}                            
}
\subfigure[]
{
\includegraphics[width=8.5cm, height=6cm]{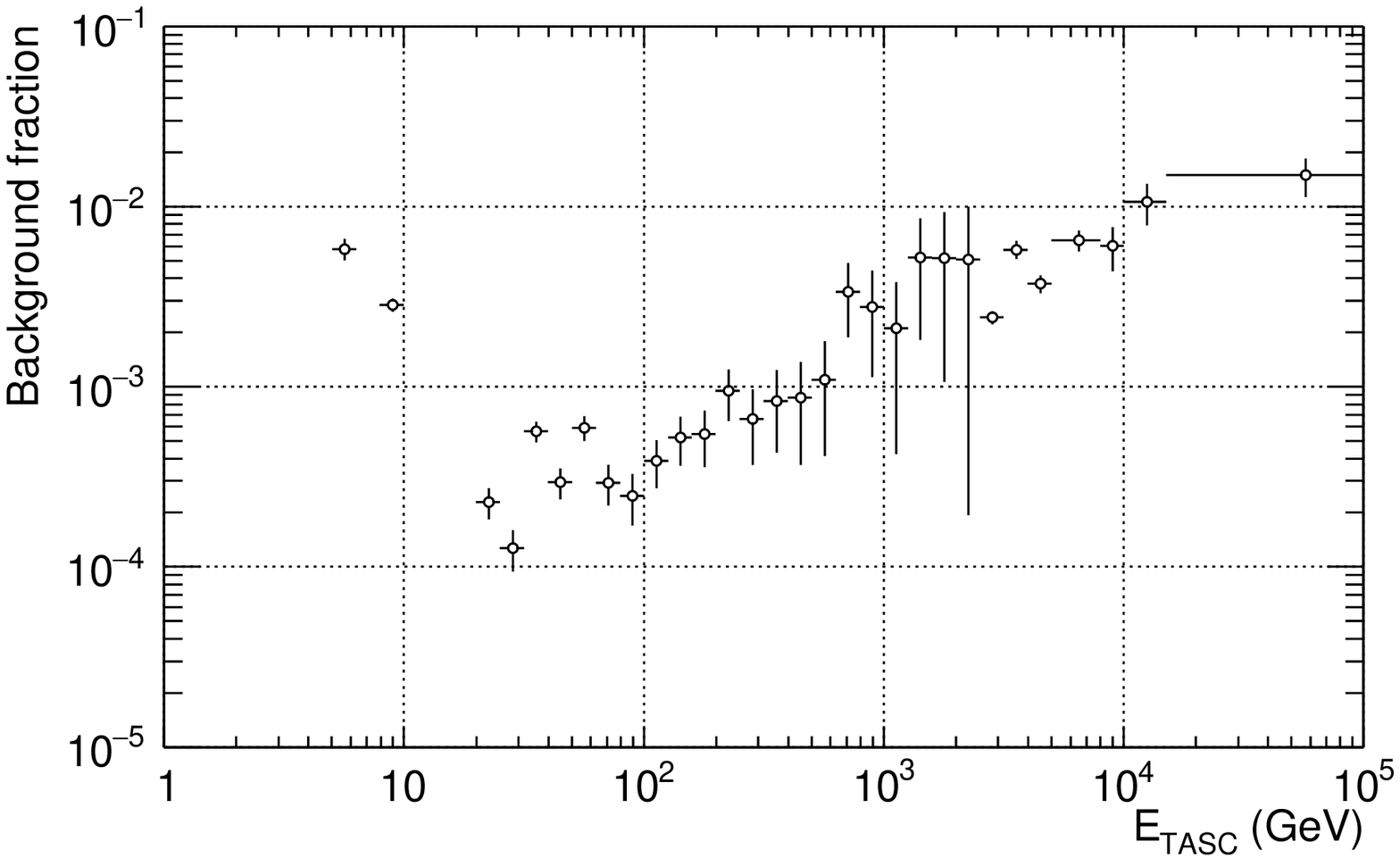}
\label{fig:bkgfracO}                            
}
\caption{Distributions of $E_{\rm TASC}$ for selected carbon (a) and oxygen (b) events in FD  (black line) and estimated contamination from different nuclei. 
The total background  (open black dots) fraction as a function of $E_{\rm TASC}$  is shown for (c) carbon and (d) oxygen.
Contamination of each nuclear species with $Z>4$  is estimated by rescaling its  $E_{TASC}$ distribution measured in FD  by the ratio, estimated with MC simulations,
of its reconstruction efficiency to the probability of being misidentified as C or O.
Background due to proton and helium is computed by normalizing their $E_{TASC}$ distributions from MC to the number of events expected 
from previous flux measurements. 
}
\label{fig:TASCedep}
\end{center}
\end{figure}
\clearpage
\section{Energy measurement}
Differently from the case of electrons, the energy released in TASC by interacting CR nuclei is only a fraction of the primary particle energy, 
the electromagnetic component of the hadronic cascades, originating from the decays of $\pi^0$ secondaries produced in the showers.
Though a significant part of the hadronic cascade energy leaks out of the calorimeter because of its limited thickness 
 (1.3 $\lambda_I$), the energy deposited in the TASC by the electromagnetic shower core scales  linearly with the incident particle energy, 
albeit with large event-to-event fluctuations. 
As a result, the energy resolution is poor by the standards of total containment hadron calorimetry in experiments at accelerators. Nevertheless, it is sufficient to reconstruct the 
steep energy spectra of CR nuclei with a nearly energy independent resolution.\\
The TASC response to nuclei was studied 
 at CERN SPS
in 2015 using a beam of accelerated ion fragments with $A/Z = 2$ and kinetic energy of 13, 19 and 150 GeV$/n$ \cite{Sakaike2015}.
In Fig.~\ref{fig:energy}a, the $E_{\rm TASC}$ distributions for C nuclei at 150 GeV/n is shown as an example. 
C nuclei in the beam are selected with CHD and the HE trigger is applied. 
The resulting distribution looks nearly gaussian, 
the energy released in the TASC is $\sim$20\% of the particle energy and the resolution $\sigma_E$ is close to 30\%. 
The mean energy deposited in TASC by different nuclear species in the beam (selected with CHD)
 is plotted as a function of the kinetic energy per particle in Fig.~\ref{fig:energy}b. 
The energy response of TASC is linear up to the maximum available particle energy of 6 TeV 
(obtained with a primary beam of $^{40}$Ar nuclei). \\ 
The energy response derived from MC simulations was tuned
using the beam test results.
Correction factors are 6.7\% for  $E_{\rm TASC}<45$ GeV and 3.5\% for $E_{\rm TASC}>350$ GeV, respectively, while
a simple linear interpolation is used to determine the correction factor for intermediate energies.\\
For flux measurement, energy unfolding is applied to correct $E_{\rm TASC}$ distributions of selected C and O candidates for significant bin-to-bin migration effects 
(due to the limited energy resolution) and infer the primary particle energy. 
In this analysis, we apply the iterative unfolding method based on the Bayes’ theorem \cite{SAgo} implemented in the RooUnfold package \cite{SROOUNFOLD,SROOT}. 
The response matrix is derived using MC simulations of the CALET flight model after applying the same selection 
as for FD and used in the unfolding procedure (Fig.~\ref{fig:SM}). Each element of the matrix represents the probability that primary nuclei in a certain 
energy interval of the CR spectrum produce an energy deposit in a given $E_{\rm TASC}$ bin.
\begin{figure}[h] \centering
\subfigure[]
{
\includegraphics[height=0.4\hsize, width=0.5\hsize]{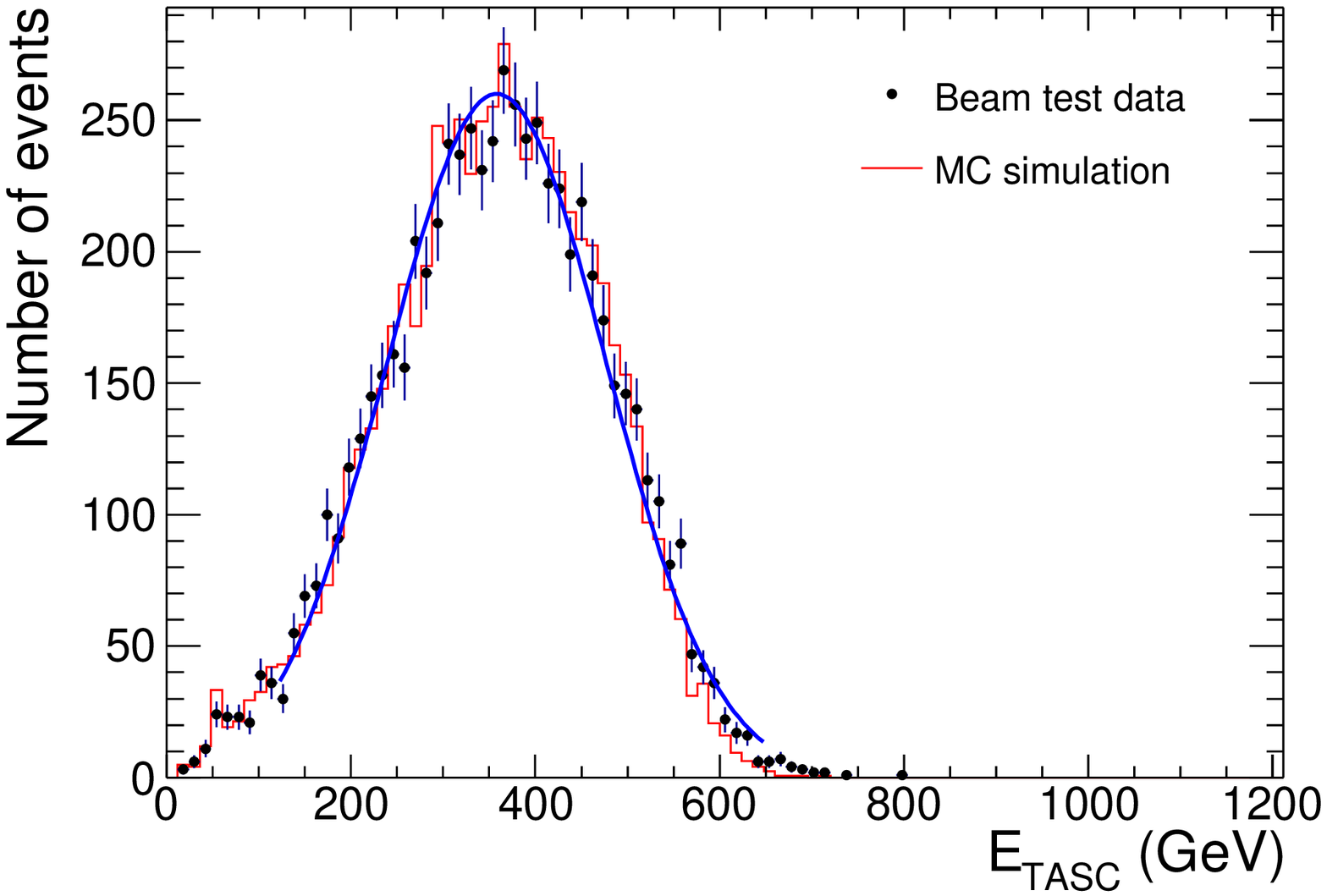}  
  \label{fig:C150tb}
}
\subfigure[]
{
\includegraphics[height=0.4\hsize]{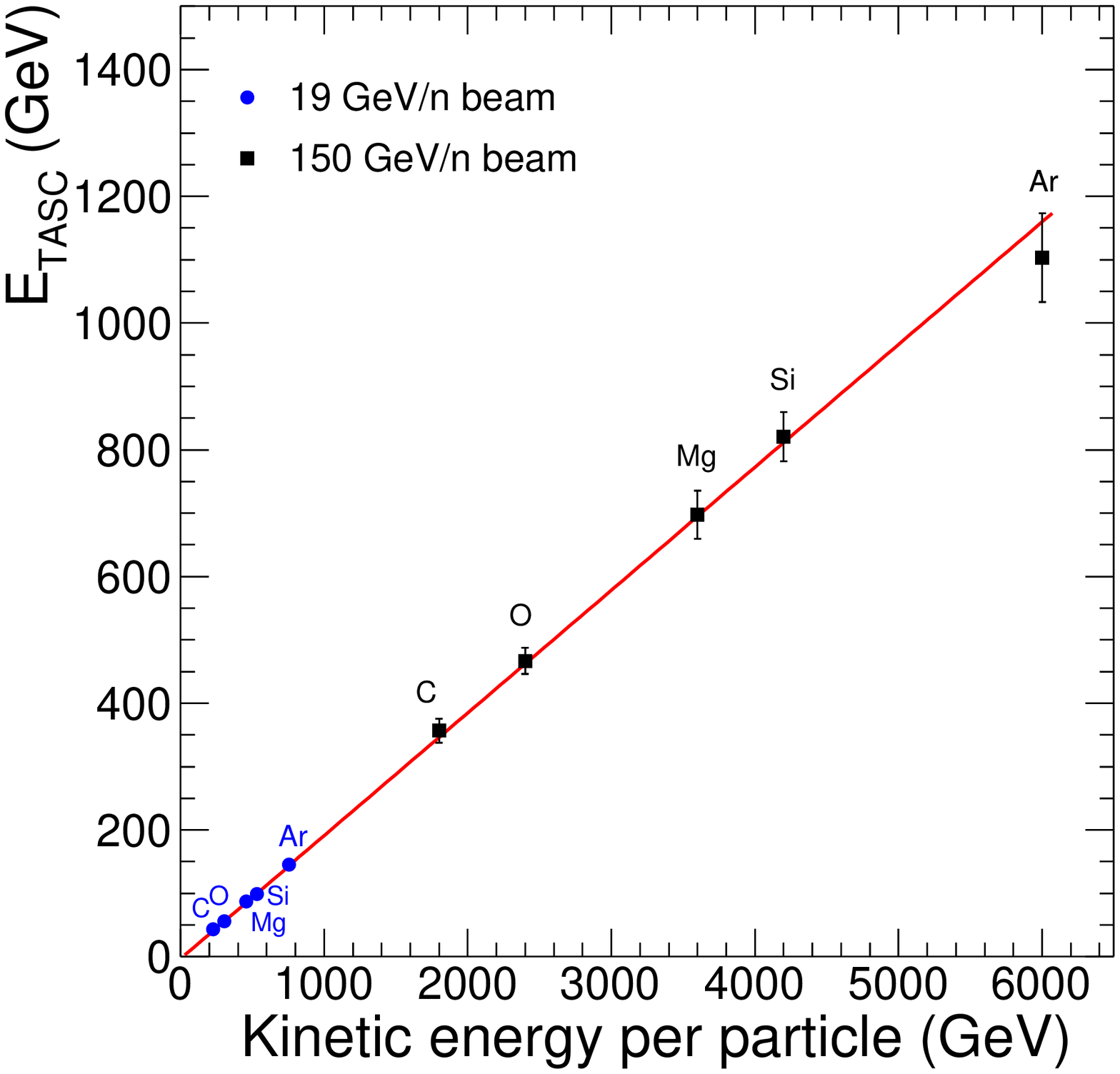} 
  \label{fig:SM}
}
\caption{ (a) Energy deposited in TASC by a beam of accelerated C nuclei of fixed energy 150 GeV/n at CERN-SPS \cite{akaike2015}.  The data are fitted to a gaussian (blue line); 
the  mean value 
is 20\% of the kinetic beam particle energy, and the energy resolution (defined as the standard deviation to mean ratio) is $\sim$30\%. 
(b)   Energy linearity of TASC as measured at CERN SPS with beams of 19 (blue dots) and 150 (black squares) GeV$/n$ ion fragments with
$A/Z = 2$. The red line represents a linear fit to the data. The fitted slope is $0.194 \pm 0.005$, indicating that on average $\sim$20\% of the particle energy is deposited in TASC. }
\label{fig:energy}
\end{figure}\noindent
\begin{figure}[hbt!] \centering
\includegraphics[width=0.5\hsize]{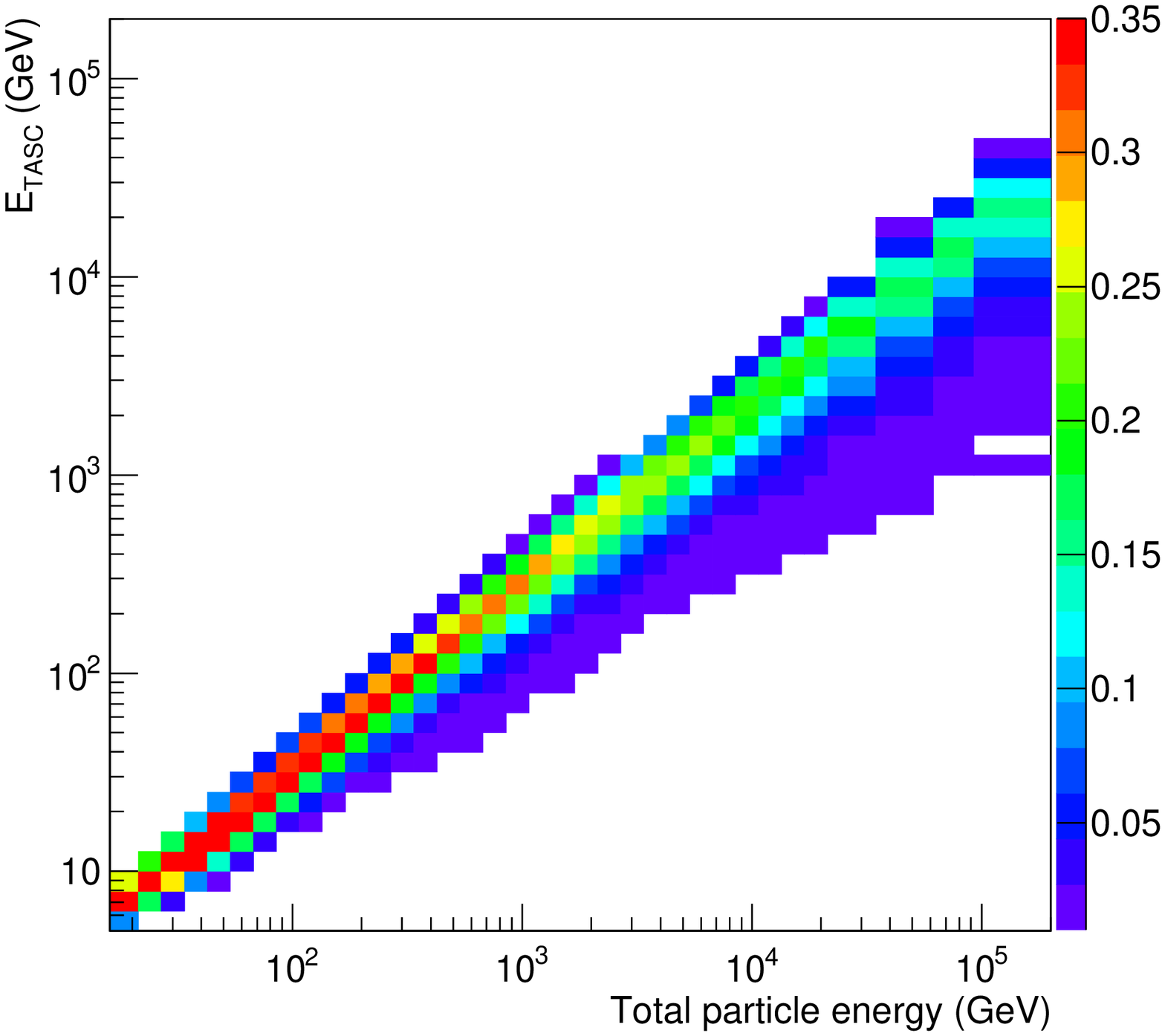} 
\caption{Response matrix for oxygen derived from MC simulations of the CALET flight model by applying the same selection as for FD. 
The color scale is associated to the probability that nuclei of a given energy produce showers in different intervals of $E_{\rm TASC}$. }
\label{fig:SM}
\end{figure}\noindent
\clearpage
\clearpage
\begin{figure}
\begin{center}
\includegraphics[scale=0.6]{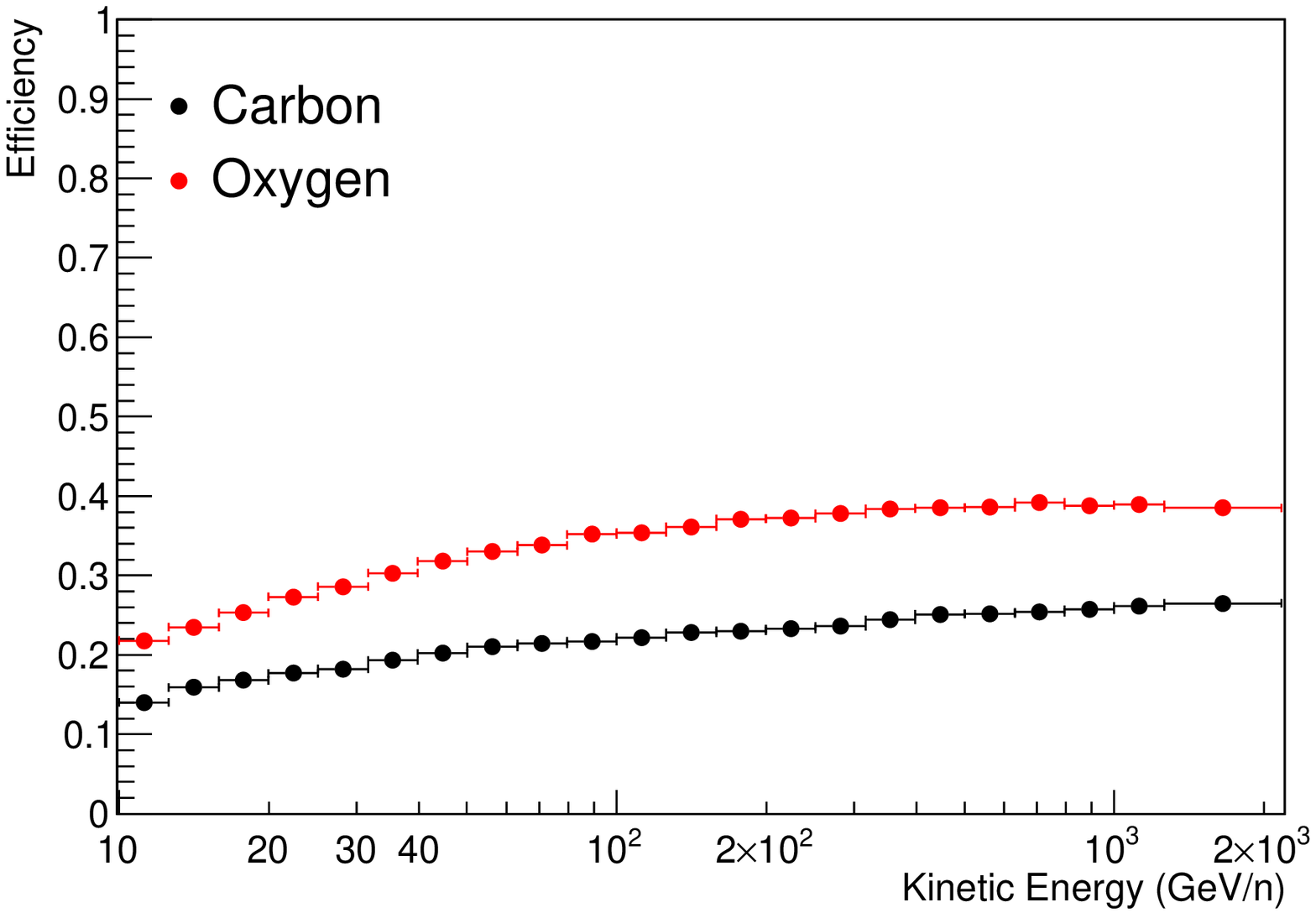}
\caption{Total selection efficiency for C (black dots) and O (red dots) estimated with MC simulations.}
\label{fig:COeff}
\end{center}
\end{figure}\noindent
\clearpage
\clearpage
\begin{figure}[hbt!] \centering
\subfigure[]
{
\includegraphics[scale=0.6]{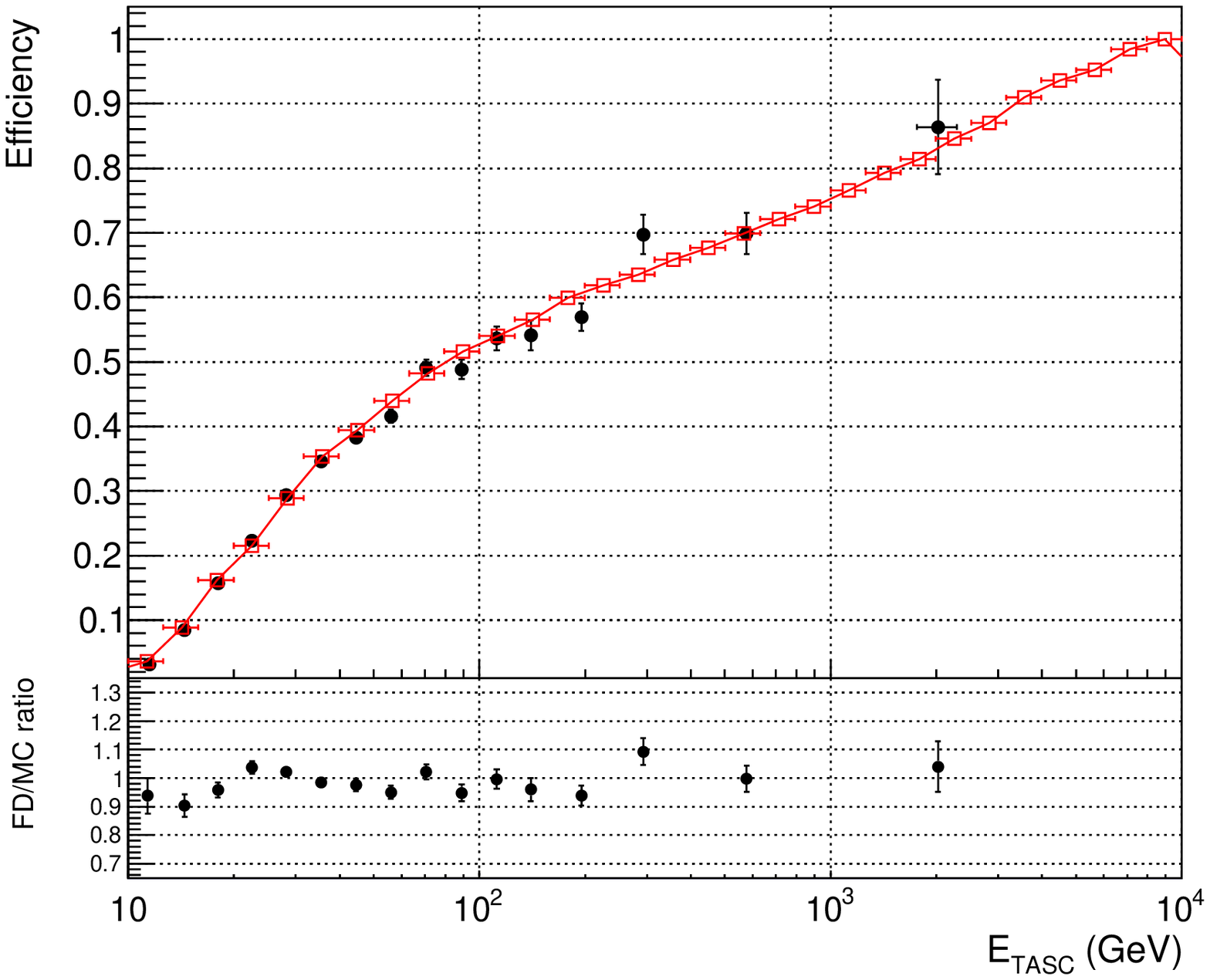}
\label{fig:HEeffC}                            
}
\subfigure[]
{
\includegraphics[scale=0.6]{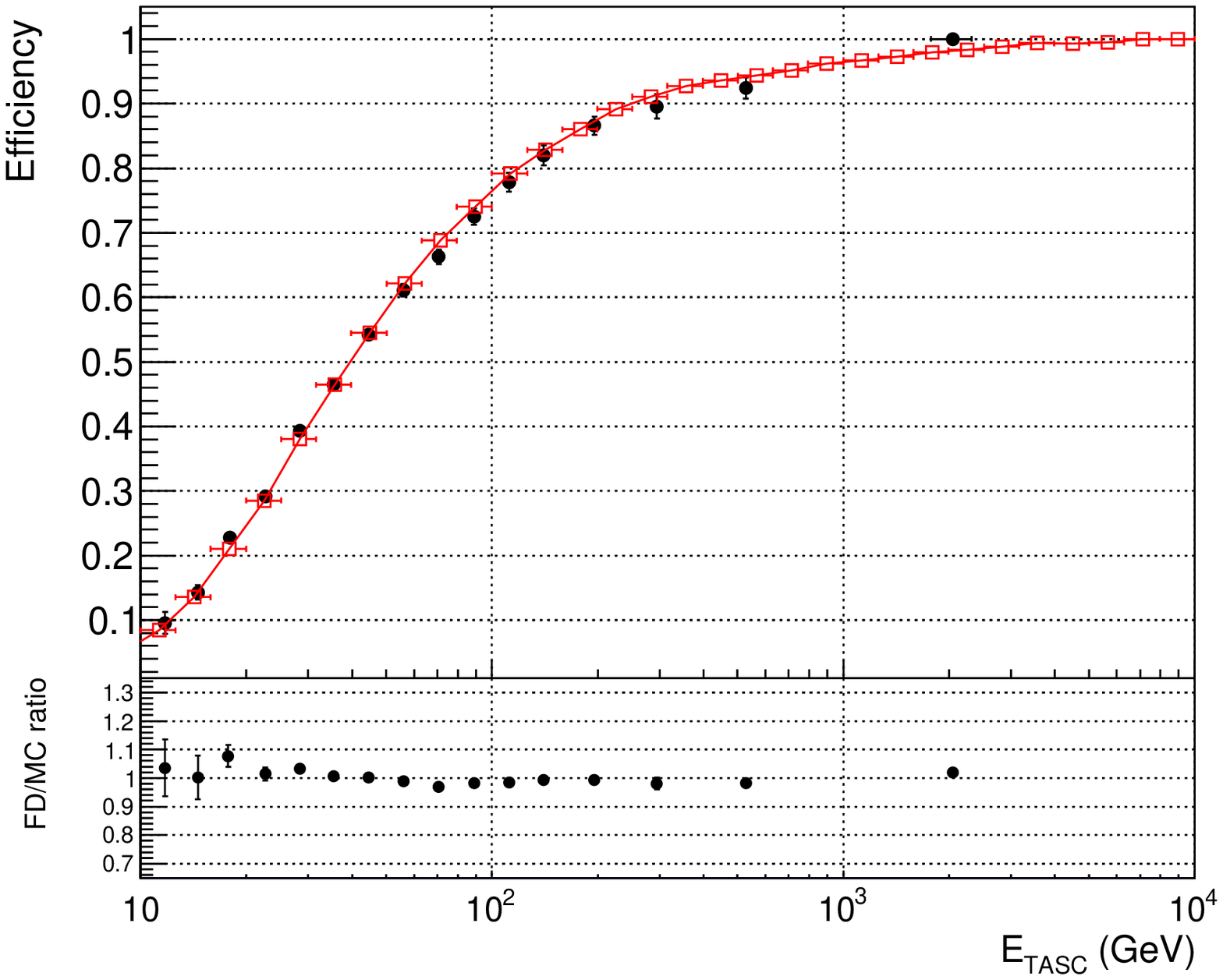}
\label{fig:HEeffO}                            
}
\caption{HE trigger efficiency as a function of the deposited energy in TASC for C ({\it a}) and O ({\it b}) as derived from flight data (FD) (black dots) and Monte Carlo (MC) (red rectangles). 
The FD to MC efficiency ratio is shown in the lower canvas.
The HE trigger efficiency was measured directly from the data by using dedicated runs where in addition to HE, a low-energy (LE) trigger was active.
The  trigger logic is the same for both trigger (i.e. coincidence of the pulse heights  of the  last two pairs of IMC layers and the top TASC layer) 
but lower discriminator thresholds are set for the input signals in case of LE trigger, 
allowing to trigger also penetrating nuclei with $Z>2$.
The ratio of the number of events counted by both  triggers to those registered  by the LE trigger only
is  an estimate  of the HE trigger efficiency in each bin of deposited energy.
To compare with simulations, we apply exactly the same method to MC samples, where both  trigger modes are modeled.
 } 
\label{fig:HEeff}
\end{figure}\noindent%
\clearpage
\section{Survival probabilities}
In order to check that hadronic interactions in the detector are well simulated, we have measured the survival probabilities 
of C and O nuclei at different depths in IMC and compared with the ones expected from MC simulations. 
For each HE-triggered event, selected as C or O by means of CHD only, 
several measurements of the charge are obtained by dE/dx samples along the particle track 
in pairs of adjacent scintillating-fiber (SciFi) layers. 
The coupled SciFi layers are spaced by about 2.3 cm and interleaved with W plates (except for the first pair), aluminum honeycomb panels and CFRP (carbon fiber reinforced polymer) supporting structures. 
From the charge distribution of each pair of Scifi layers, the number of events that did not interact in the material upstream the fibers
can be estimated by selecting the nuclei with a charge value consistent with the one measured in CHD. 
The reduction in the number of not-interacted events, normalized to the ones selected in CHD, allows to measure the survival probability as a function 
of the material thickness traversed by the particle in the IMC, as shown in Fig.~\ref{fig:survprob}.
In MC data, the survival probabilities can be calculated by using the true information on the point where
the first hadronic interaction occurs in the detector. 
The survival probabilities measured in FD are in good agreement (within $<1\%$) with MC predictions, as shown in the bottom panels of Fig.~\ref{fig:survprob}.
\begin{figure}[hbt!] \centering
\subfigure[]
{
  \includegraphics[width=0.48\hsize]{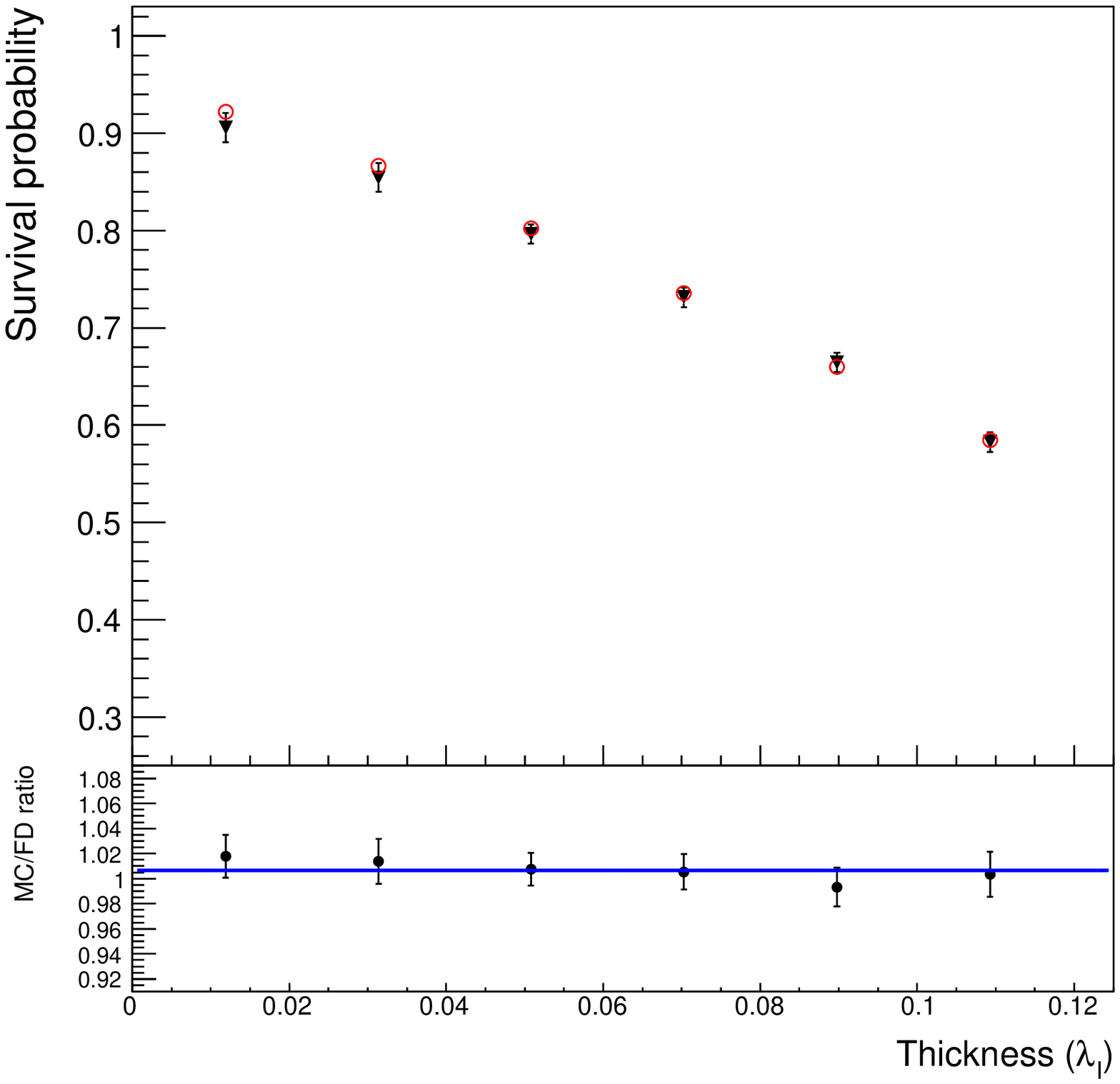}
  \label{fig:survprobC}                            
}
\subfigure[]
{
  \includegraphics[width=0.48\hsize]{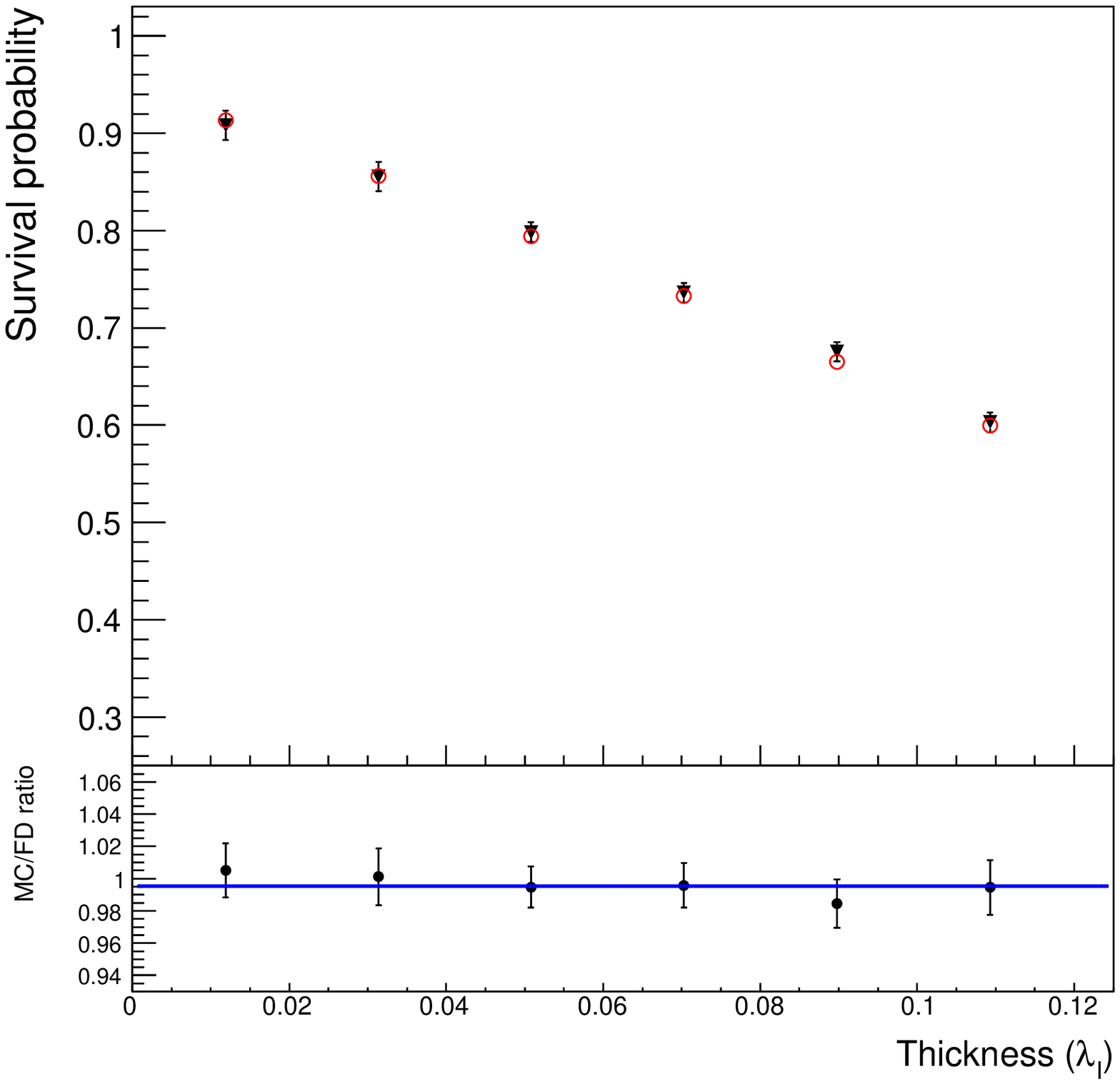}
  \label{fig:survprobO}                            
}
\caption{Survival probability  as a function of  the material thickness traversed by the particles in IMC
as derived from FD (black triangles) and MC (red circles) for carbon ({\it a}) and oxygen ({\it b}) (upper panels).  The survival probabilities are calculated in FD  by dividing the number of events 
selected as C (O) in the first six pairs of SciFi layers in IMC by the number of C (O) events selected with CHD. 
The material thickness is expressed in units of proton interaction length $\lambda_I$ and it is measured from the bottom of CHD. 
The coupled SciFi layers are preceded by 0.2 $X_0$-thick W plates (except the first pair), Al honeycomb panels and CFRP supporting structures. 
In bottom panels, the blue dotted line represents a constant value fitted to the ratio between MC and FD survival probabilities. 
The fitted value is $0.997\pm0.008$ ($1.006\pm0.006$) for O (C).
}
\label{fig:survprob}
\end{figure}
\noindent
\clearpage
\begin{figure}[hbt!] \centering
\subfigure[]
{
  \includegraphics[width=15cm, height=9cm]{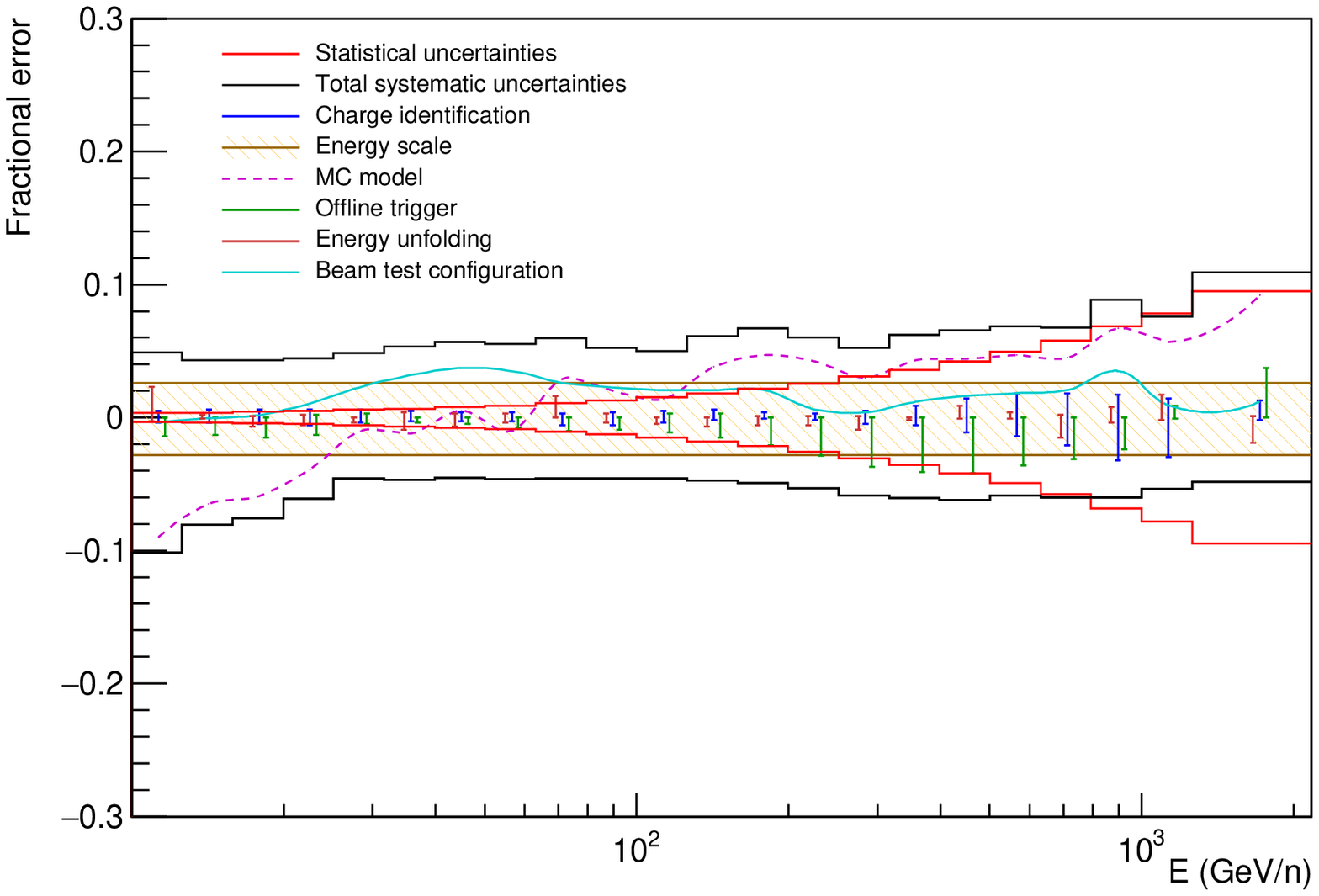}
  \label{fig:sysC}                            
}
\subfigure[]
{
  \includegraphics[width=15cm, height=9cm]{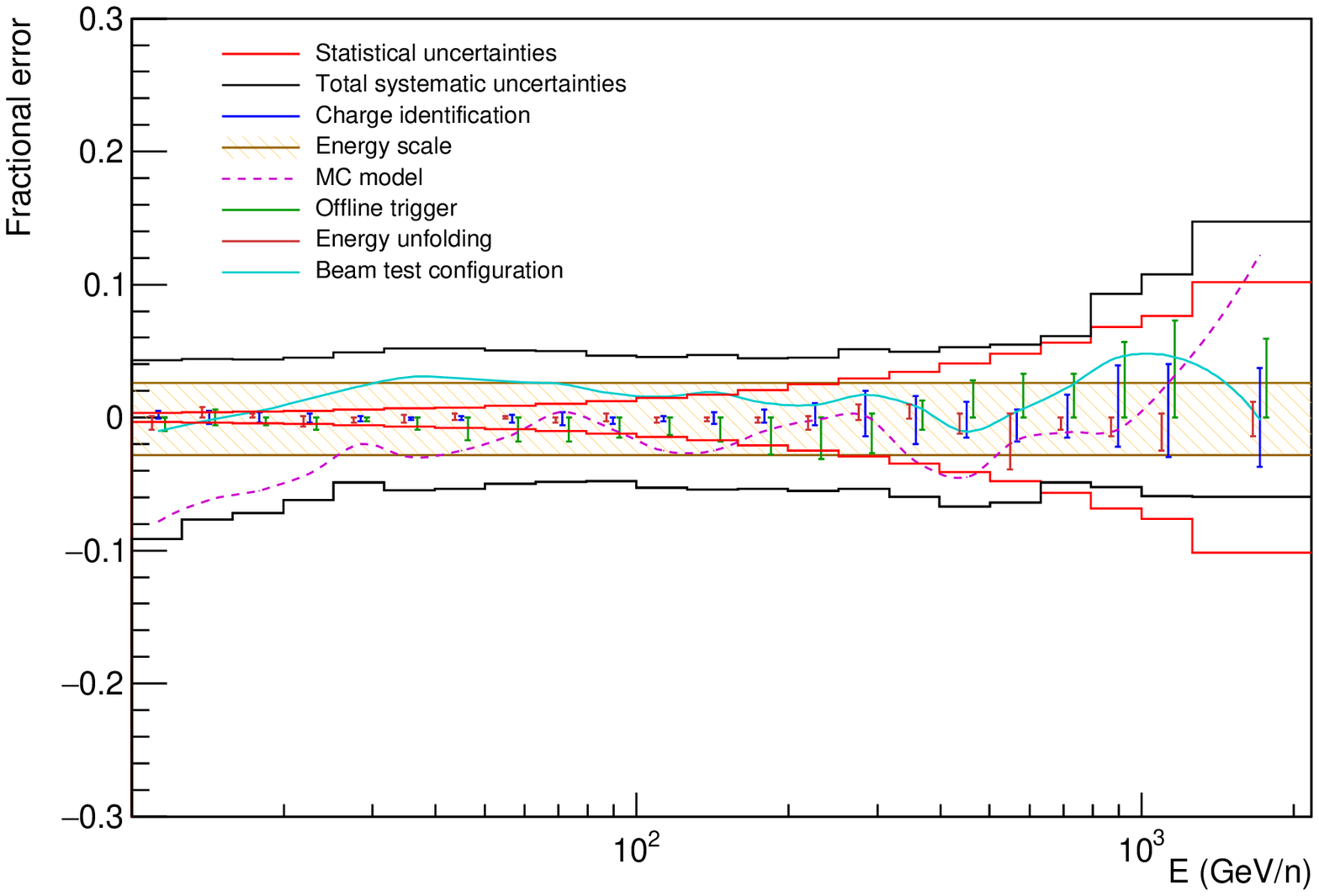}
  \label{fig:sysO}                            
}
\caption{Energy (per nucleon in GeV) dependence of  systematic uncertainties (relative errors) for C  ({\it a}) and O ({\it b}).  The band defined by the red lines represents the statistical error in each energy bin. The band within the black lines shows the sum in quadrature of all the sources of systematics.  A detailed breakdown of systematic errors, stemming from charge identification, offline trigger, MC model, energy scale correction, energy unfolding and beam test configuration, is shown.
}
\label{fig:sys_all}
\end{figure}\noindent
\clearpage
\begin{figure}[hbt!] \centering
\includegraphics[width=0.7\hsize]{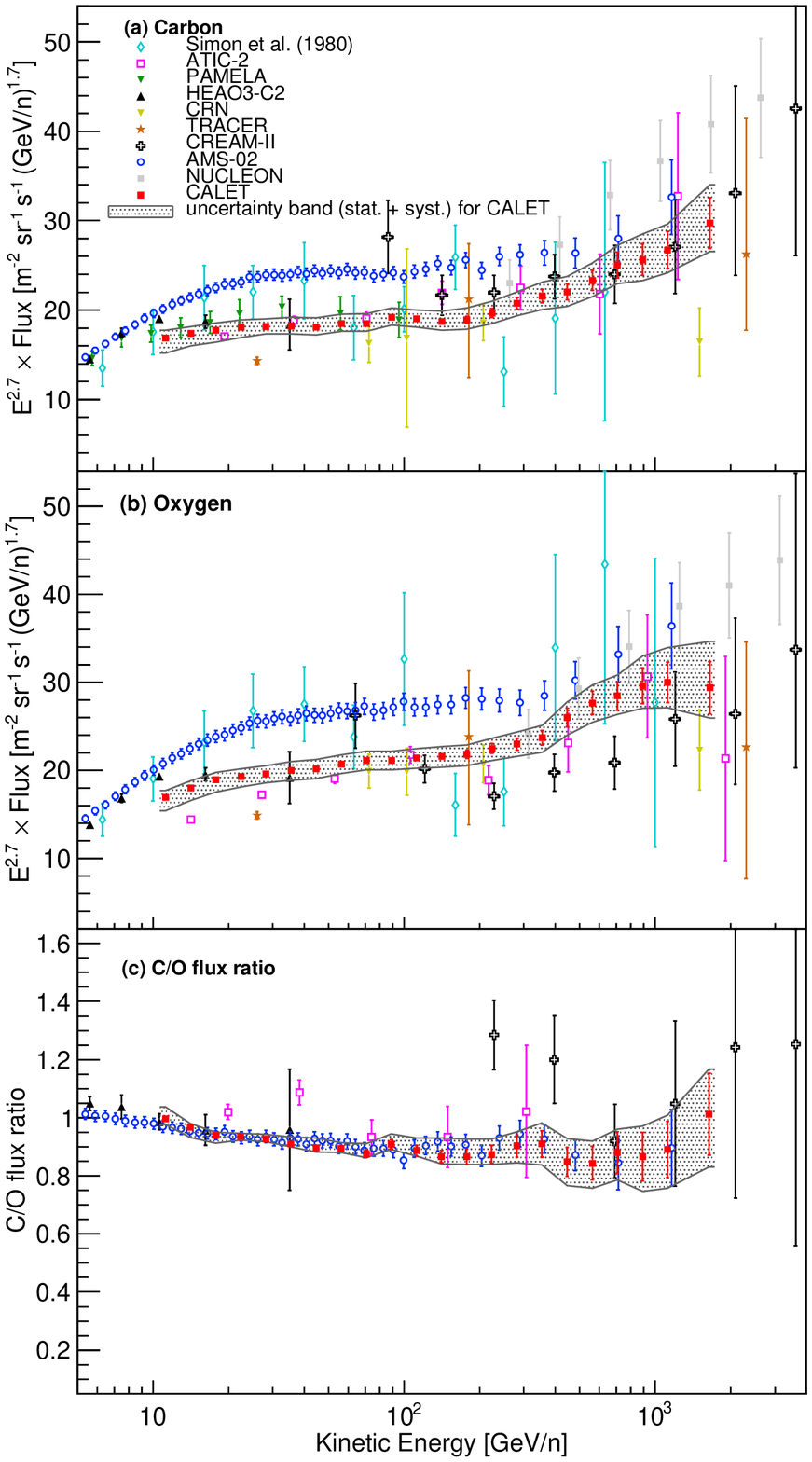} 
\caption{CALET (a) carbon and (b) oxygen flux (multiplied by $E^{2.7}$) and (c)  ratio of carbon to oxygen fluxes, as a function of kinetic energy $E$.
Error bars of CALET data (red) represent the statistical uncertainty only, while the  gray band indicates the quadratic sum of statistical and systematic errors. 
Also plotted are other direct measurements \cite{SHEAO, SCRN, SATIC2, STRACER, SAMS-CO, SPAMELA-C, SCREAM2, SNUCLEON, SSimon}.}
\label{fig:flux}
\end{figure}\noindent
\clearpage
\begin{figure}
\begin{center}
\subfigure[]
{
\includegraphics[width=0.8\hsize]{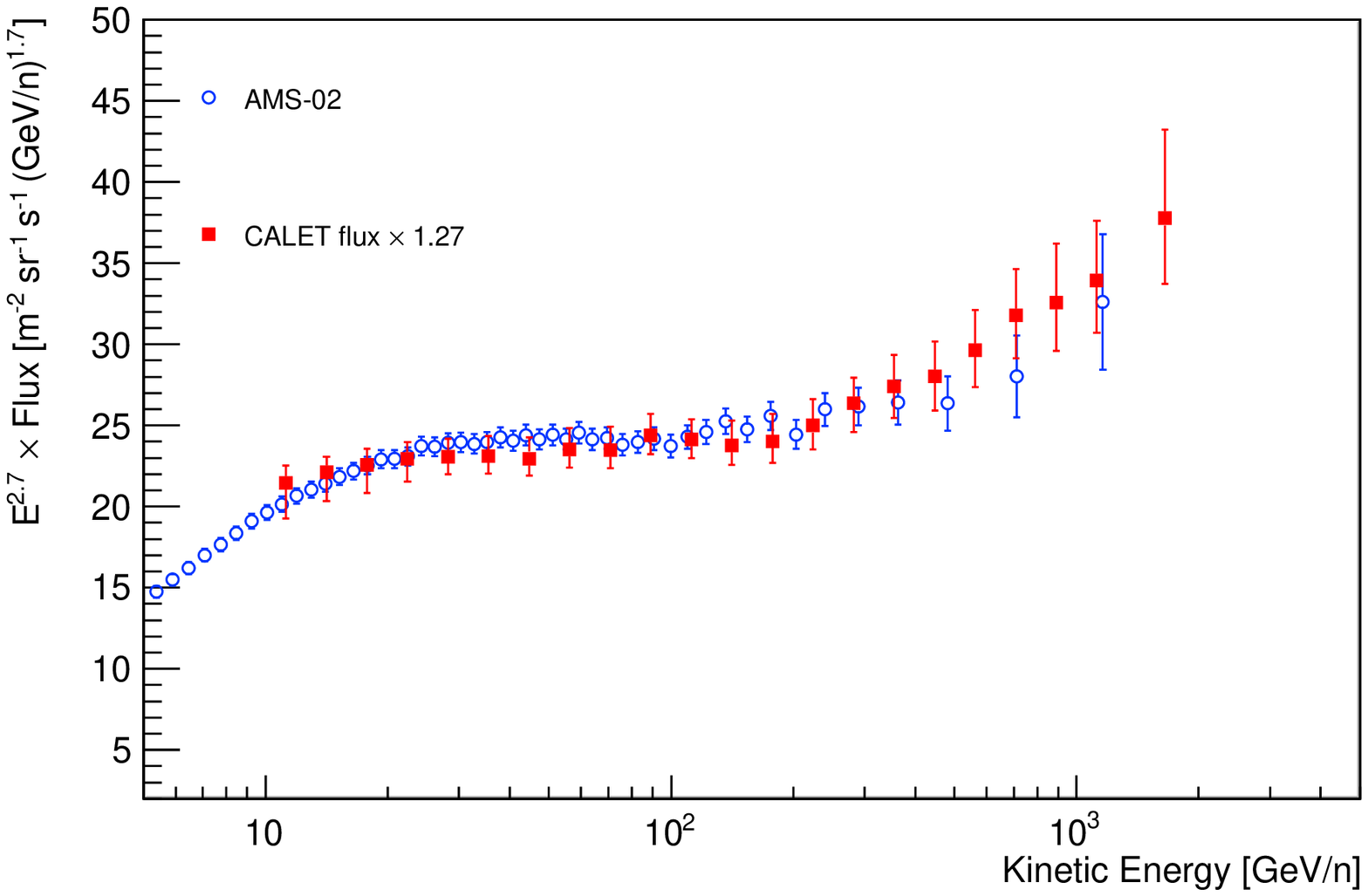}
\label{fig:fluxC1.28}                            
}
\subfigure[]
{
\includegraphics[width=0.8\hsize]{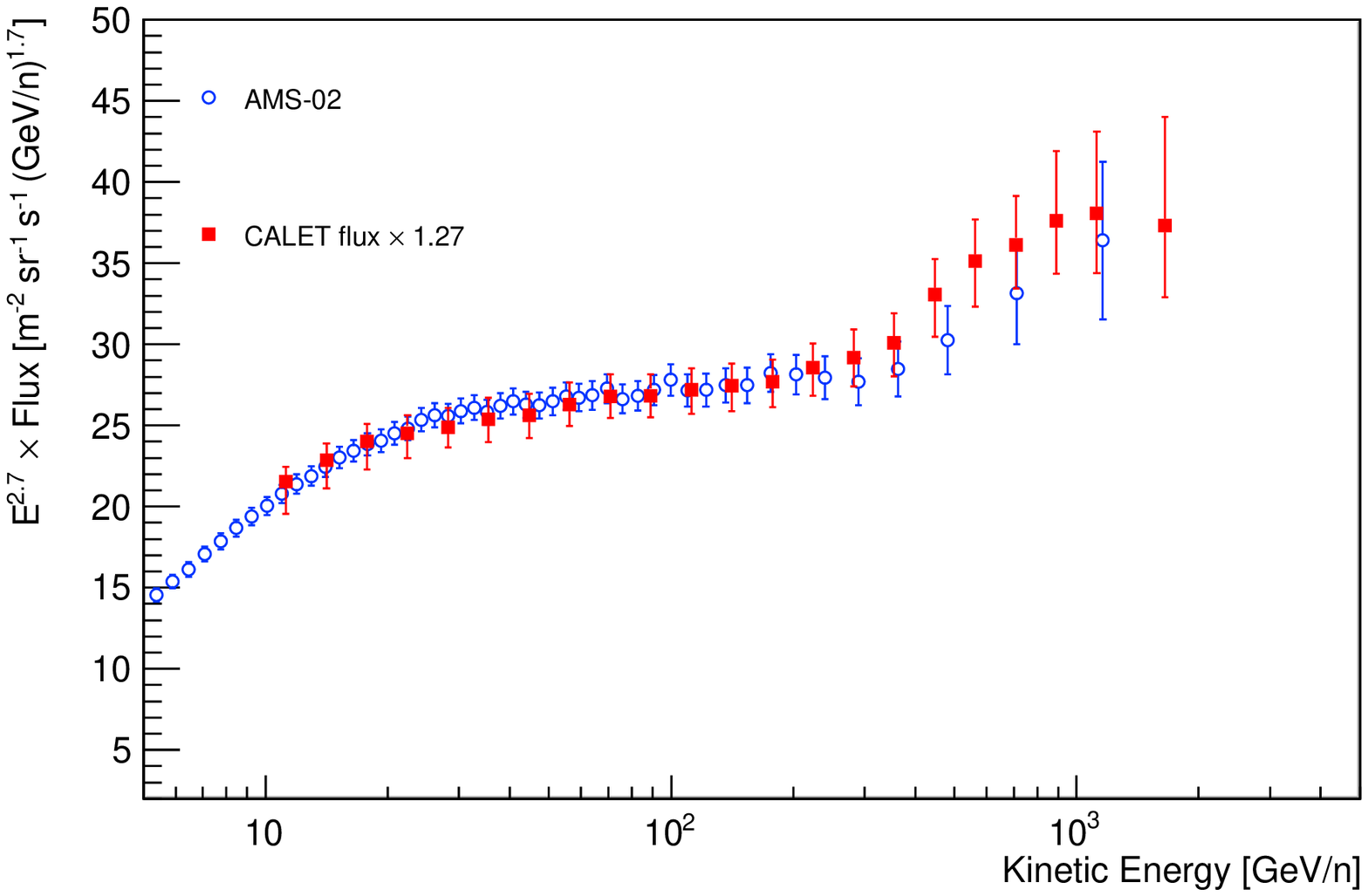}
\label{fig:fluxO1.28}                            
}
\caption{({\it a}) Carbon and ({\it b}) oxygen fluxes measured by CALET (red points) are multiplied by 1.27 for comparison with AMS-02 results \cite{SAMS-CO}.
Error bars of CALET data represent the quadratic sum of statistical and systematic uncertainties.
The factor (1.27$\pm$0.03)  is obtained by a simultaneous minimization of the sum of squared residuals between CALET and AMS data points for C and O. 
}
\label{fig:flux1.28}
\end{center}
\end{figure}
\clearpage
\section{Spectral fit method}
We have tested two models for the C and O energy spectra: 
a double power-law (DPL) in energy 
\begin{equation}
\Phi(E) = \begin{cases} C \left(\frac{E}{\text{GeV}} \right)^{\gamma} & E\le E_0\\
C \left(\frac{E}{\text{GeV}} \right)^{\gamma}  \left(\frac{E}{E_0}\right)^{\Delta\gamma}   & E>E_0 \end{cases}
\label{eq:DPL}
\end{equation}
and a single power-law (SPL)
\begin{equation}
\Phi(E) = C\, \left(\frac{E}{\text{GeV}} \right)^{\gamma}
\label{eq:SPL}
\end{equation}
where $C$ is a normalization factor, $\gamma$ the spectral index, and 
$\Delta \gamma$ the spectral index change above the transition energy $E_0$.  
We have fitted the data using a chi-square minimization. The effect of systematic uncertainties in the measurement of the energy spectrum is modeled 
introducing in the $\chi^2$ function a set of nuisance parameters. \\
The  $\chi^2$ function  is defined as
\begin{equation}
\chi^2 = \sum_{i=4}^{N} \left[ \frac{\Phi_i + S(E_i, \boldsymbol{\beta} ) \sigma_{i}^{sys}  - \Phi(E_i, \bf{p})}{\sigma^{stat}_i} \right]^2 + \sum_{j=1}^{N_{np}} \beta_i^2
\label{eq:chi2}
\end{equation}
where $\Phi_i$, $\sigma_i^{stat}$, $\sigma_i^{sys}$, $E_i$ are  the measured flux, the statistical and 
systematic errors, and the energy of the $i$-th bin (geometric mean of bin edges), respectively; 
$N = 22$ is the number of data points and the fit starts from the fourth point. 
The elements of the vector $\bf{p}$ are the free parameters of the model function $\Phi$ used in the fit:
${\bf p} = \{C, \gamma, \Delta\gamma, E_0\}$ for Eq.~\ref{eq:DPL}, and 
${\bf p} = \{C, \gamma\}$ for Eq.~\ref{eq:SPL}.\\
$\boldsymbol{\beta}$ is a set of independent nuisance parameters $\beta_j$ with $j=1,..., N_{np}$
introduced to properly account  for systematic uncertainties in the fit.
The energy range to be fitted (from 25 to 2200 GeV$/n$) is logarithmically divided into  $N_{np}$ intervals.
According to the energy dependence of the systematic fractional errors (Fig.~\ref{fig:sys_all}), we chose  $N_{np}=6$;
therefore the width of each interval covers three consecutive energy bins of the spectrum.
A nuisance parameter  $\beta_j$ is assigned to each interval. 
In the fit, a gaussian constraint (the last term in Eq.~\ref{eq:chi2}) is applied to each parameter  $\beta_j$, with zero mean and standard deviation equal to 1. 
In the $\chi^2$ function, the systematic errors $\sigma_i^{sys}$ are multiplied by a  piece-wise function $S(E_i, \boldsymbol{\beta})$,  
which assumes the value of $\beta_j$ of the corresponding  energy interval within which $E_i$ falls.\\
The results of the fits with the two models are shown in Figs.~\ref{fig:Cfit_sys} and \ref{fig:Ofit_sys} for C and O spectrum, respectively.
We also performed the fits by varying the number of nuisance parameters, with similar results and significance. 
\clearpage
\begin{figure} \centering
\includegraphics[width=0.8\hsize]{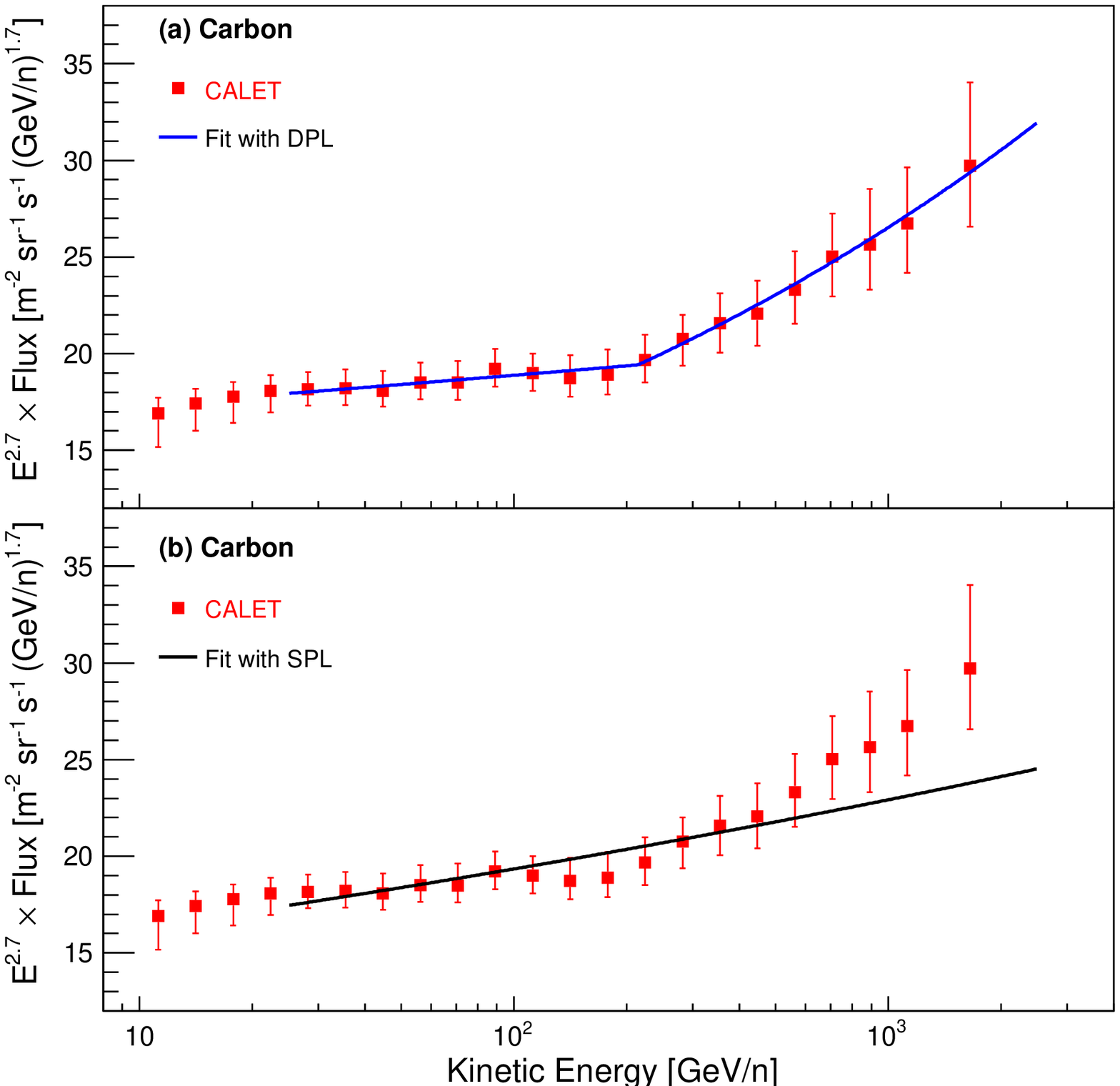}
\caption{Fit to the carbon energy spectrum with (a) a DPL function (Eq.~\ref{eq:DPL}) and (b) a SPL function (Eq.~\ref{eq:SPL}), in the energy range [25, 2000] GeV/n. 
Error bars of CALET data points represent the sum in quadrature of statistical and systematic uncertainties. 
The DPL fit yields  $\gamma = -2.663\pm0.014$,   $E_0 = (215\pm54)$ GeV$/n$, $\Delta\gamma = 0.166\pm0.042$, with $\chi^2/$d.o.f. = 9.0/8.
The SPL fit yields $\gamma = -2.626\pm0.010$ with $\chi^2/$d.o.f. = 27.5/10.
The difference $\Delta\chi^2=18.5$ between the fits with the two models, 
 with two additional free parameters in DPL fit with respect to SPL fit, 
allows to exclude the single power law hypothesis at the 3.9$\sigma$ level.}
\label{fig:Cfit_sys}
\end{figure}\noindent
\clearpage\begin{figure} \centering
\includegraphics[width=0.8\hsize]{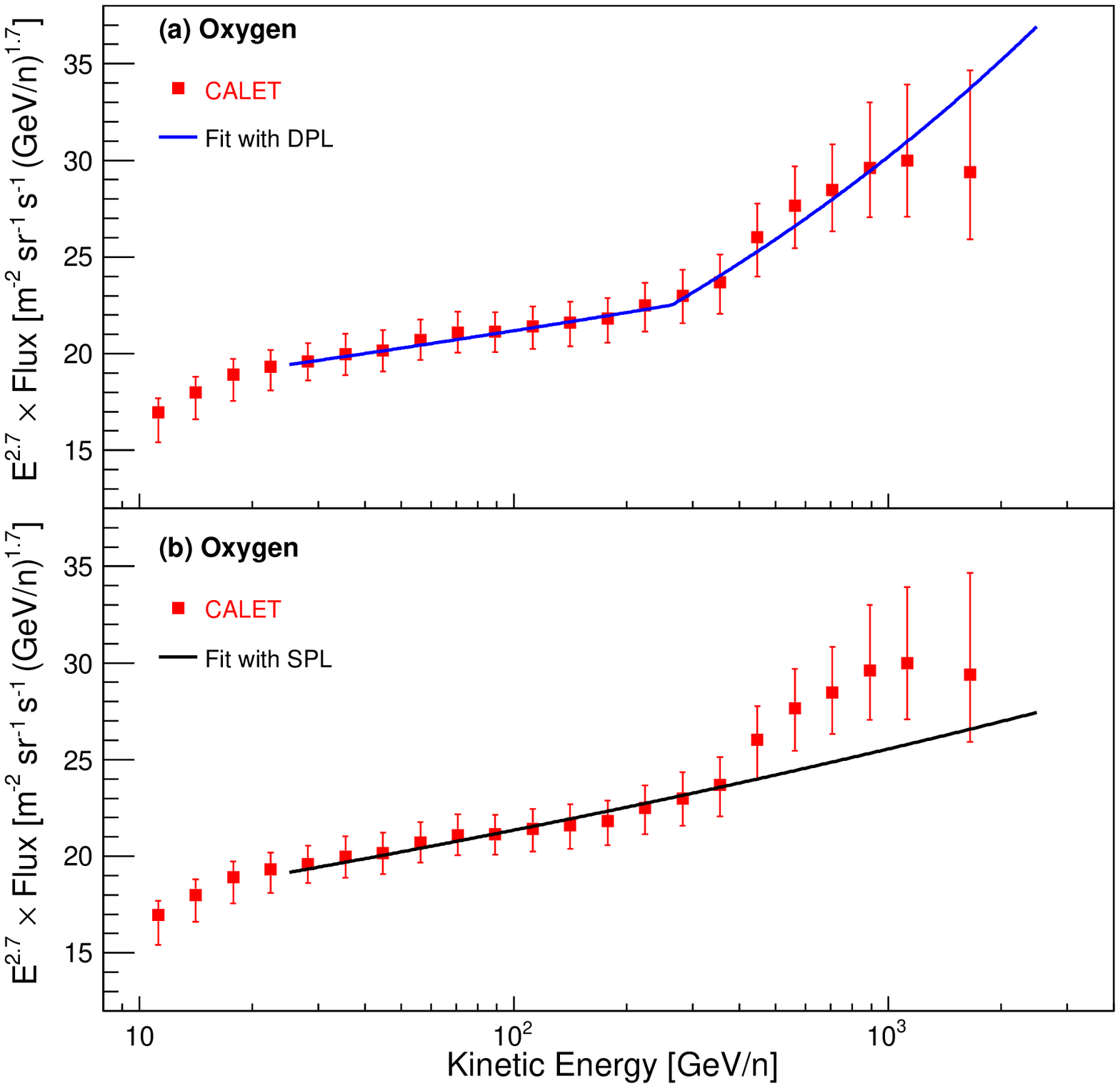}
\caption{Fit to the oxygen energy spectrum with (a) a DPL function (Eq.~\ref{eq:DPL}) and (b) a SPL function (Eq.~\ref{eq:SPL}), in the energy range [25, 2000]  GeV$/n$. 
Error bars of CALET data points represent the sum in quadrature of statistical and systematic uncertainties. 
The DPL  fit yields $\gamma = -2.637\pm0.009$,  $E_0 = (264\pm53)$ GeV$/n$, $\Delta\gamma = 0.158\pm0.053$, with $\chi^2/$d.o.f. = 3.0/8.
The SPL fit yields $\gamma=-2.622\pm0.008$ with $\chi^2/$d.o.f. = 15.9/10.
The difference $\Delta\chi^2=12.9$ between the fits with the two models, 
 with two additional free parameters in DPL fit with respect to SPL fit, 
allows to exclude the single power law hypothesis at the 3.2$\sigma$ level.}
\label{fig:Ofit_sys}
\end{figure}\noindent
\clearpage\begin{figure} \centering
\includegraphics[width=0.8\hsize]{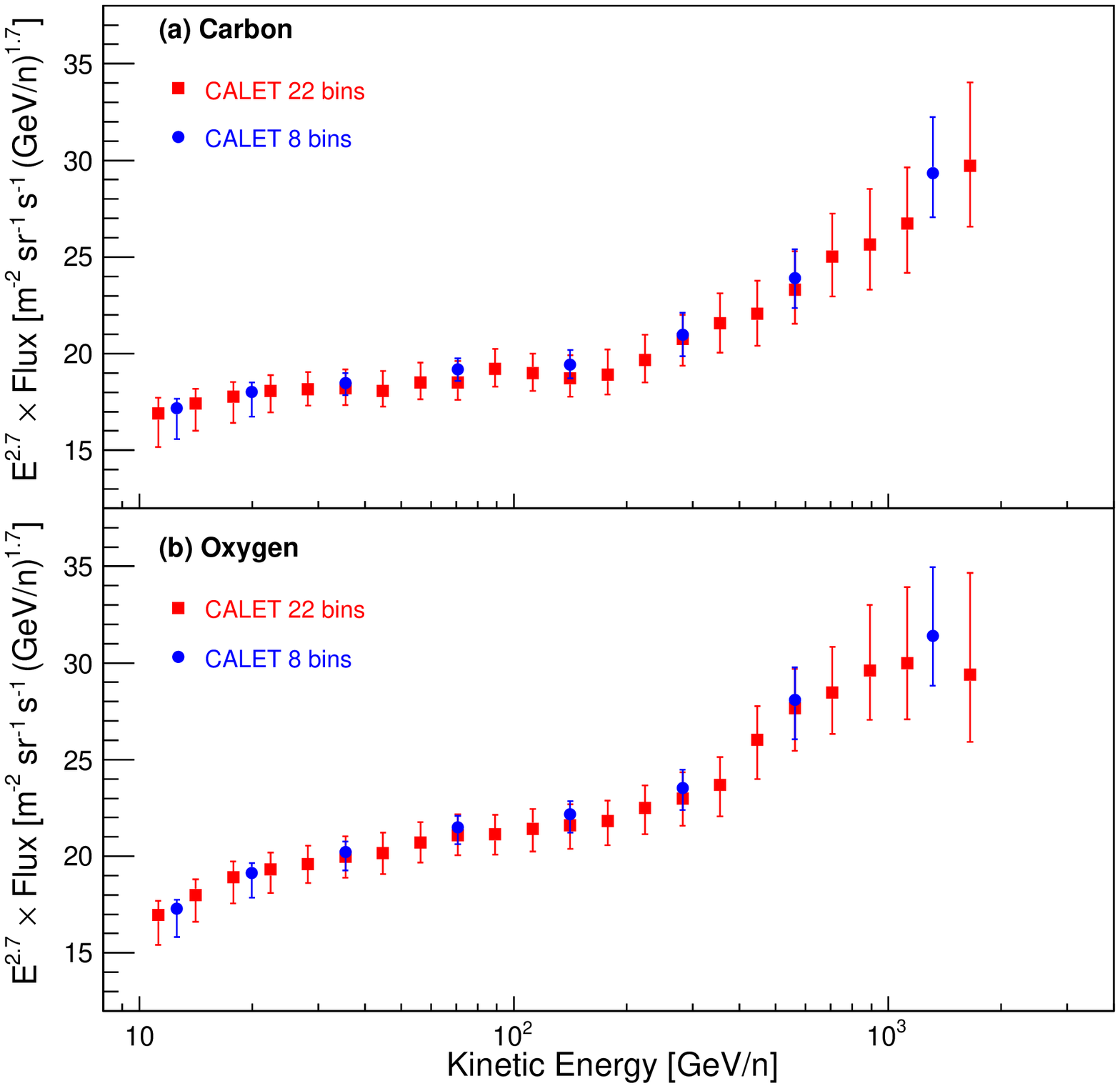}
\caption{\textcolor{black}{CALET carbon ({\it a})  and oxygen  ({\it b}) spectra derived with two different energy binnings. 
In the reference spectra (red squares) the energy range from 10 GeV$/n$ to 2.2 TeV$/n$  is divided into 22 bins (Tables I and II); 
all bins are chosen to have relative width commensurate with the TASC energy resolution $\sigma_E$, with the exception of the last bin whose width is $2\times\sigma_E$.
To study possible binning related effects in the spectra, C and O fluxes are also  derived by dividing the energy range into 8 bins (blue dots), with 
relative bin width $\sim2.5\times\sigma_E$, but the last bin whose width is $3.5\times\sigma_E$.
The wide-bin spectra have the same shape as the original ones. 
The flux differences are well within the error bars, representing the quadratic sum of statistical and systematic uncertainties. }
}
\label{fig:CObinning}
\end{figure}\noindent
\clearpage
\begin{figure} \centering
\includegraphics[width=0.8\hsize]{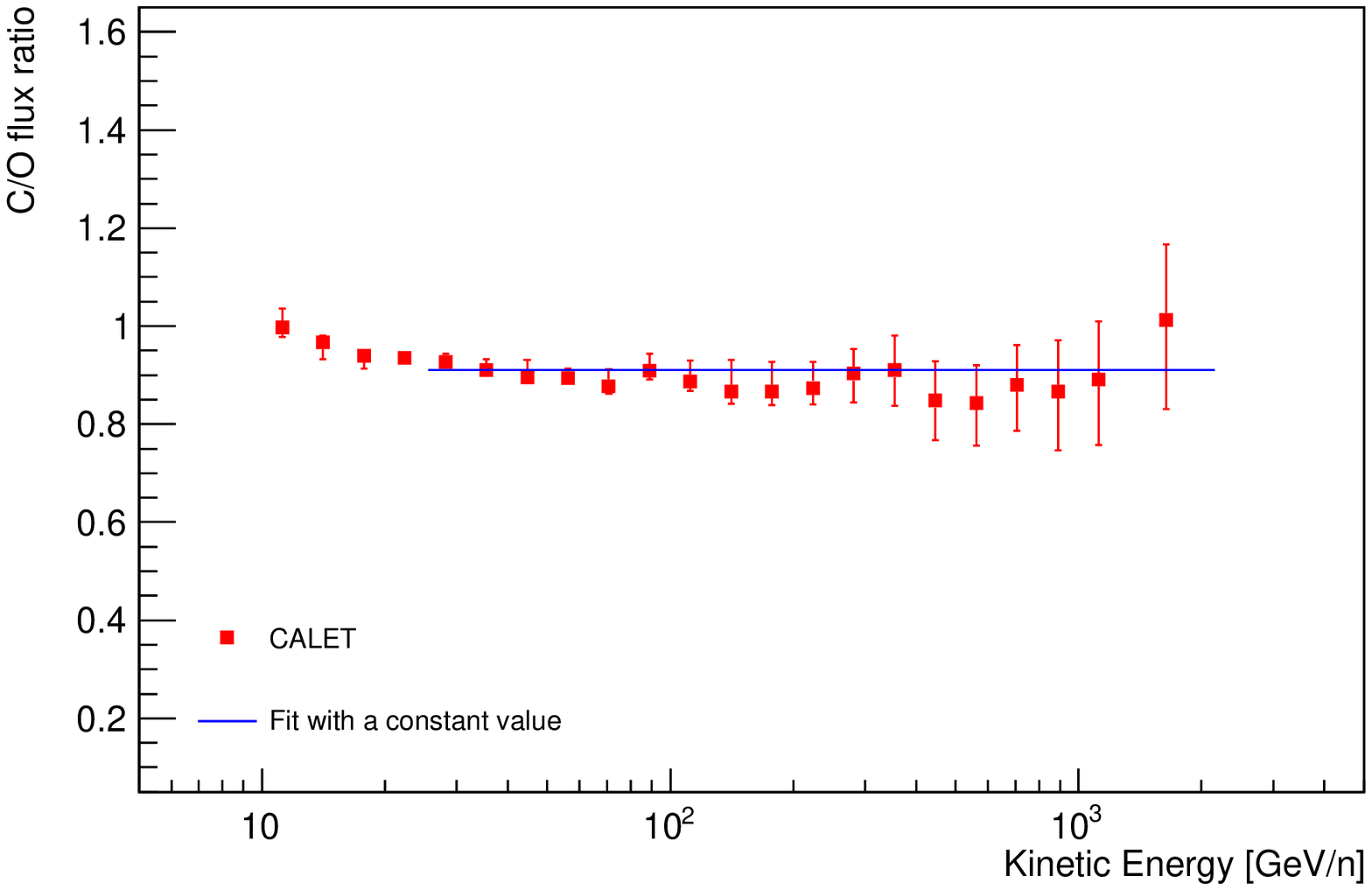}
\caption{The carbon to oxygen flux ratio as a function of kinetic energy per nucleon is fitted to a constant function (blue line). 
Error bars of CALET data points represent the sum in quadrature of statistical and systematic uncertainties. 
Above 25 GeV$/n$ the C/O ratio is described by a constant value of $0.911\pm0.006$ with $\chi^2/$d.o.f. = 8.3/17.
}
\label{fig:COratiofit}
\end{figure}\noindent
\clearpage
\renewcommand{\arraystretch}{1.25}
\begin{table*}
  \caption{Table of CALET carbon spectrum. 
 The first, second and third error in the flux represents the statistical uncertainties, systematic uncertainties in normalization, and energy dependent systematic uncertainties, respectively.
\label{tab:Cflux}}
\begin{ruledtabular}
\begin{tabular}{c c c c}
Energy Bin [GeV$/n$] & Flux [m$^{-2}$sr$^{-1}$s$^{-1}$(GeV$/n$)$^{-1}$]   \\
\hline
  10.0--12.6 & $(2.471 \, \pm 0.008 \, _{-0.110}^{+0.106} \, _{-0.226}^{+0.058}) \times 10^{-2}$ \\ 
  12.6--15.8 & $(1.368 \, \pm 0.005 \, _{-0.061}^{+0.059} \, _{-0.092}^{+0.009}) \times 10^{-2}$ \\ 
  15.8--20.0 & $(7.492 \, \pm 0.032 \, _{-0.332}^{+0.321} \, _{-0.461}^{+0.047}) \times 10^{-3}$ \\ 
  20.0--25.1 & $(4.092 \, \pm 0.020 \, _{-0.182}^{+0.175} \, _{-0.172}^{+0.051}) \times 10^{-3}$ \\ 
  25.1--31.6 & $(2.208 \, \pm 0.013 \, _{-0.098}^{+0.095} \, _{-0.028}^{+0.051}) \times 10^{-3}$ \\ 
  31.6--39.8 & $(1.189 \, \pm 0.008 \, _{-0.053}^{+0.051} \, _{-0.019}^{+0.038}) \times 10^{-3}$ \\ 
  39.8--50.1 & $(6.341 \, \pm 0.049 \, _{-0.281}^{+0.271} \, _{-0.058}^{+0.236}) \times 10^{-4}$ \\ 
  50.1--63.1 & $(3.487 \, \pm 0.031 \, _{-0.155}^{+0.149} \, _{-0.048}^{+0.122}) \times 10^{-4}$ \\ 
  63.1--79.4 & $(1.871 \, \pm 0.020 \, _{-0.083}^{+0.080} \, _{-0.022}^{+0.078}) \times 10^{-4}$ \\ 
  79.4--100.0 & $(1.044 \, \pm 0.013 \, _{-0.046}^{+0.045} \, _{-0.012}^{+0.031}) \times 10^{-4}$ \\ 
 100.0--125.9 & $(5.545 \, \pm 0.084 \, _{-0.246}^{+0.237} \, _{-0.071}^{+0.141}) \times 10^{-5}$ \\ 
 125.9--158.5 & $(2.932 \, \pm 0.053 \, _{-0.130}^{+0.125} \, _{-0.049}^{+0.128}) \times 10^{-5}$ \\ 
 158.5--199.5 & $(1.590 \, \pm 0.034 \, _{-0.071}^{+0.068} \, _{-0.035}^{+0.082}) \times 10^{-5}$ \\ 
 199.5--251.2 & $(8.890 \, \pm 0.228 \, _{-0.394}^{+0.381} \, _{-0.264}^{+0.378}) \times 10^{-6}$ \\ 
 251.2--316.2 & $(5.035 \, \pm 0.154 \, _{-0.223}^{+0.216} \, _{-0.193}^{+0.154}) \times 10^{-6}$ \\ 
 316.2--398.1 & $(2.812 \, \pm 0.100 \, _{-0.125}^{+0.120} \, _{-0.117}^{+0.127}) \times 10^{-6}$ \\ 
 398.1--501.2 & $(1.544 \, \pm 0.065 \, _{-0.068}^{+0.066} \, _{-0.067}^{+0.076}) \times 10^{-6}$ \\ 
 501.2--631.0 & $(8.766 \, \pm 0.432 \, _{-0.389}^{+0.375} \, _{-0.339}^{+0.468}) \times 10^{-7}$ \\ 
 631.0--794.3 & $(5.053 \, \pm 0.292 \, _{-0.224}^{+0.216} \, _{-0.204}^{+0.264}) \times 10^{-7}$ \\ 
 794.3--1000.0 & $(2.780 \, \pm 0.190 \, _{-0.123}^{+0.119} \, _{-0.112}^{+0.215}) \times 10^{-7}$ \\ 
1000.0--1258.9 & $(1.555 \, \pm 0.122 \, _{-0.069}^{+0.067} \, _{-0.047}^{+0.097}) \times 10^{-7}$ \\ 
1258.9--2166.7 & $(6.094 \, \pm 0.578 \, _{-0.270}^{+0.261} \, _{-0.116}^{+0.613}) \times 10^{-8}$ \\ 
\end{tabular}
\end{ruledtabular}
\end{table*}
\renewcommand{\arraystretch}{1.0}

\clearpage
\renewcommand{\arraystretch}{1.25}
\begin{table*}
  \caption{Table of CALET oxygen spectrum. 
  The first, second and third error in the flux represents the statistical uncertainties, systematic uncertainties in normalization, and energy dependent systematic uncertainties, respectively.
\label{tab:Oflux}}
\begin{ruledtabular}
\begin{tabular}{c c c}
Energy Bin [GeV$/n$] & Flux [m$^{-2}$sr$^{-1}$s$^{-1}$(GeV$/n$)$^{-1}$]   \\
\hline
 10.0--12.6 & $(2.479 \, \pm 0.009 \, _{-0.110}^{+0.106} \, _{-0.198}^{+0.013}) \times 10^{-2}$ \\ 
  12.6--15.8 & $(1.414 \, \pm 0.005 \, _{-0.063}^{+0.061} \, _{-0.088}^{+0.016}) \times 10^{-2}$ \\ 
  15.8--20.0 & $(7.977 \, \pm 0.035 \, _{-0.354}^{+0.341} \, _{-0.450}^{+0.056}) \times 10^{-3}$ \\ 
  20.0--25.1 & $(4.374 \, \pm 0.022 \, _{-0.194}^{+0.187} \, _{-0.191}^{+0.063}) \times 10^{-3}$ \\ 
  25.1--31.6 & $(2.381 \, \pm 0.014 \, _{-0.106}^{+0.102} \, _{-0.050}^{+0.056}) \times 10^{-3}$ \\ 
  31.6--39.8 & $(1.305 \, \pm 0.009 \, _{-0.058}^{+0.056} \, _{-0.041}^{+0.039}) \times 10^{-3}$ \\ 
  39.8--50.1 & $(7.074 \, \pm 0.054 \, _{-0.314}^{+0.303} \, _{-0.215}^{+0.210}) \times 10^{-4}$ \\ 
  50.1--63.1 & $(3.901 \, \pm 0.035 \, _{-0.173}^{+0.167} \, _{-0.088}^{+0.106}) \times 10^{-4}$ \\ 
  63.1--79.4 & $(2.134 \, \pm 0.022 \, _{-0.095}^{+0.091} \, _{-0.041}^{+0.055}) \times 10^{-4}$ \\ 
  79.4--100.0 & $(1.148 \, \pm 0.014 \, _{-0.051}^{+0.049} \, _{-0.021}^{+0.021}) \times 10^{-4}$ \\ 
 100.0--125.9 & $(6.250 \, \pm 0.092 \, _{-0.277}^{+0.267} \, _{-0.179}^{+0.097}) \times 10^{-5}$ \\ 
 125.9--158.5 & $(3.384 \, \pm 0.058 \, _{-0.150}^{+0.145} \, _{-0.106}^{+0.066}) \times 10^{-5}$ \\ 
 158.5--199.5 & $(1.836 \, \pm 0.038 \, _{-0.081}^{+0.079} \, _{-0.056}^{+0.023}) \times 10^{-5}$ \\ 
 199.5--251.2 & $(1.017 \, \pm 0.025 \, _{-0.045}^{+0.044} \, _{-0.033}^{+0.015}) \times 10^{-5}$ \\ 
 251.2--316.2 & $(5.577 \, \pm 0.164 \, _{-0.247}^{+0.239} \, _{-0.170}^{+0.158}) \times 10^{-6}$ \\ 
 316.2--398.1 & $(3.088 \, \pm 0.106 \, _{-0.137}^{+0.132} \, _{-0.122}^{+0.075}) \times 10^{-6}$ \\ 
 398.1--501.2 & $(1.821 \, \pm 0.074 \, _{-0.081}^{+0.078} \, _{-0.091}^{+0.056}) \times 10^{-6}$ \\ 
 501.2--631.0 & $(1.040 \, \pm 0.050 \, _{-0.046}^{+0.045} \, _{-0.048}^{+0.035}) \times 10^{-6}$ \\ 
 631.0--794.3 & $(5.744 \, \pm 0.324 \, _{-0.255}^{+0.246} \, _{-0.119}^{+0.249}) \times 10^{-7}$ \\ 
 794.3--1000.0 & $(3.209 \, \pm 0.219 \, _{-0.142}^{+0.137} \, _{-0.089}^{+0.265}) \times 10^{-7}$ \\ 
1000.0--1258.9 & $(1.745 \, \pm 0.133 \, _{-0.077}^{+0.075} \, _{-0.068}^{+0.173}) \times 10^{-7}$ \\ 
1258.9--2166.7 & $(6.022 \, \pm 0.613 \, _{-0.267}^{+0.258} \, _{-0.238}^{+0.849}) \times 10^{-8}$ \\ 
\end{tabular}
\end{ruledtabular}
\end{table*}
\renewcommand{\arraystretch}{1.0}

\clearpage
\renewcommand{\arraystretch}{1.25}
\begin{table*}
  \caption{Table of CALET carbon to oxygen flux ratio. 
  The first and second  error represents the statistical uncertainties and systematic uncertainties, respectively.
\label{tab:COratio}}
\begin{ruledtabular}
\begin{tabular}{c c c}
Energy Bin [GeV$/n$] & C/O    \\
\hline
 10.0--12.6 & $ 0.997 \, \pm 0.005 \, _{-0.019}^{+0.039}$  \\ 
  12.6--15.8 & $ 0.967 \, \pm 0.005 \, _{-0.034}^{+0.013}$  \\ 
  15.8--20.0 & $ 0.939 \, \pm 0.006 \, _{-0.026}^{+0.010}$  \\ 
  20.0--25.1 & $ 0.935 \, \pm 0.006 \, _{-0.010}^{+0.009}$  \\ 
  25.1--31.6 & $ 0.927 \, \pm 0.008 \, _{-0.006}^{+0.014}$  \\ 
  31.6--39.8 & $ 0.911 \, \pm 0.009 \, _{-0.006}^{+0.020}$  \\ 
  39.8--50.1 & $ 0.896 \, \pm 0.010 \, _{-0.009}^{+0.033}$  \\ 
  50.1--63.1 & $ 0.894 \, \pm 0.011 \, _{-0.007}^{+0.016}$  \\ 
  63.1--79.4 & $ 0.877 \, \pm 0.013 \, _{-0.008}^{+0.032}$  \\ 
  79.4--100.0 & $ 0.909 \, \pm 0.016 \, _{-0.008}^{+0.031}$  \\ 
 100.0--125.9 & $ 0.887 \, \pm 0.019 \, _{-0.004}^{+0.038}$  \\ 
 125.9--158.5 & $ 0.866 \, \pm 0.022 \, _{-0.010}^{+0.061}$  \\ 
 158.5--199.5 & $ 0.866 \, \pm 0.026 \, _{-0.008}^{+0.055}$  \\ 
 199.5--251.2 & $ 0.874 \, \pm 0.031 \, _{-0.014}^{+0.043}$  \\ 
 251.2--316.2 & $ 0.903 \, \pm 0.038 \, _{-0.045}^{+0.031}$  \\ 
 316.2--398.1 & $ 0.910 \, \pm 0.045 \, _{-0.056}^{+0.055}$  \\ 
 398.1--501.2 & $ 0.848 \, \pm 0.050 \, _{-0.064}^{+0.064}$  \\ 
 501.2--631.0 & $ 0.843 \, \pm 0.058 \, _{-0.064}^{+0.051}$  \\ 
 631.0--794.3 & $ 0.880 \, \pm 0.071 \, _{-0.061}^{+0.039}$  \\ 
 794.3--1000.0 & $ 0.866 \, \pm 0.084 \, _{-0.085}^{+0.063}$  \\ 
1000.0--1258.9 & $ 0.891 \, \pm 0.097 \, _{-0.092}^{+0.068}$  \\ 
1258.9--2166.7 & $ 1.012 \, \pm 0.141 \, _{-0.115}^{+0.065}$  \\ 
\hline
\end{tabular}
\end{ruledtabular}
\end{table*}
\renewcommand{\arraystretch}{1.0}

\providecommand{\noopsort}[1]{}\providecommand{\singleletter}[1]{#1}%

\end{document}